\documentclass[10pt]{article}

\usepackage{graphicx}% Include figure files
\usepackage{graphics}% Include figure files
\usepackage{epsfig}

\newcommand {\Bd} {\ensuremath{B^0}}
\newcommand {\Bs} {\ensuremath{B^0_s}}
\newcommand {\Bsbar} {\ensuremath{\bar B^0_s}}
\newcommand {\Bcpm} {\ensuremath{B_c^\pm}}
\newcommand {\Bcm} {\ensuremath{B_c^-}}
\newcommand {\Bcp} {\ensuremath{B_c^+}}
\newcommand {\Bpm} {\ensuremath{B^\pm}}
\newcommand {\Bp} {\ensuremath{B^+}}
\newcommand {\Bm} {\ensuremath{B^-}}

\newcommand {\Lb} {\ensuremath{\Lambda^0_b}}
\newcommand {\barLb} {\ensuremath{\bar{\Lambda}^0_b}}
\newcommand {\Xibpm} {\ensuremath{\Xi^\pm_b}}
\newcommand {\Xibm} {\ensuremath{\Xi^-_b}}
\newcommand {\Xibn} {\ensuremath{\Xi^0_b}}
\newcommand {\Ombpm} {\ensuremath{\Omega^\pm_b}}
\newcommand {\Ombm}  {\ensuremath{\Omega^-_b}}

\newcommand {\barBd} {\ensuremath{\bar{B}^0_d}}
\newcommand {\barBs} {\ensuremath{\bar{B}^0_s}}

\newcommand {\asld} {\ensuremath{a^d_{\mathrm{sl}}}}
\newcommand {\asls} {\ensuremath{a^s_{\mathrm{sl}}}}
\newcommand {\aslq} {\ensuremath{a^q_{\mathrm{sl}}}}
\newcommand {\aslb} {\ensuremath{A^b_{\mathrm{sl}}}}

\newcommand {\Ks} {\ensuremath{K_S^0}}

\begin{document}

\markboth{G.Borissov}
{$B$-physics Results from Tevatron}

%\catchline{}{}{}{}{}
%
%%%%%%%%%%%%%%%%%%%%%%%%%%%%%%%%%%%%%%%%%%%%%%%%%%%%%%%%%%%%%%%%%%%%

%\rightline{FERMILAB preprint}
%\rightline{ \mbox{Fermilab-Pub-04/xxx-E}}

%\title{{\small \rightline{test}} \\
\title{\mbox{\small \rightline{Fermilab-Pub-13-091-PPD}} \\ 
[1.2ex]
$B$-PHYSICS RESULTS FROM TEVATRON}

%\author{GUENNADI BORISSOV}

\author{Guennadi Borissov
\thanks{To be published in International Journal of Modern Physics A}\\
\small Department of Physics, Lancaster University, \\[-0.8ex]
\small Lancaster, LA1 4YB, United Kingdom\\
\small \texttt{g.borissov@lancaster.ac.uk}\\
}

\date{4 January 2013}
%\address{
%Department of Physics, Lancaster University \\
%Lancaster, LA1 4YB, United Kingdom \\
%g.borissov@lancaster.ac.uk
%}

\maketitle

%\begin{history}
%\received{Day Month Year}
%\revised{Day Month Year}
%\end{history}

\begin{abstract}
This review summarizes the most important results in $B$ physics
obtained at the Tevatron. They include the discovery of the new
$B$ hadrons, the measurement of their masses and lifetimes, the
measurement of the oscillation frequency of $\Bs$ meson, the search
for its rare decay $\Bs \to \mu^+ \mu^-$, and the study of the $CP$
asymmetry in decays and mixing of $B$ mesons. \\
[1.2ex]

%\keywords{$B$ physics; $CP$ asymmetry; $\Bs$ mixing; Rare decays.}

%\ccode{PACS numbers:}

\end{abstract}

\section{Introduction}
\label{intro}

The experiments CDF and D\O\ collected more than 10 fb$^{-1}$ of $p \bar{p}$ collisions
in RunII of the Tevatron collider at Fermilab from 2002 to 2011
at a center of mass energy $\sqrt{s} = 1.96$ TeV.
%They worked at center of mass energy $\sqrt{s} = 1.96$ TeV,
It was the world highest energy during almost the entire period of data taking,
which allowed the experiments to search
for new phenomena at the energy frontier of the particle physics. Namely this search
at the highest possible collision energy and in particular hunting for new heavy particles
%which would not be possible elsewhere,
was the main goal of both experiments. They were designed accordingly as the multipurpose
detectors capable to detect and study the processes with the center of mass energy of the order of 1 TeV.

However, both experiments in the same time actively pursued the study of $B$ hadrons. This direction
may look unusual and unexpected for the Tevatron, as the high energy of $p \bar p$ collisions is not needed,
while the high multiplicity and the overwhelming
background coming from light quark and gluon interactions are clearly obstructive for this kind of research.
What is required, on the contrary, is a dedicated experimental facility
with such special features as sophisticated trigger on $B$-hadron production, exclusive reconstruction
of all $B$-hadron decay products, both charged and neutral,
precise measurement of their trajectories and identification of their type.
These requirements were only partially satisfied for the CDF and D\O\ detectors. On the contrary,
the specialised experiments at $B$ factories, such as BaBar and Belle,
were operational at the same time as the Tevatron experiments, possessed all these
qualities and ran in a much cleaner environment of $e^+ e^-$ collisions.
And it would seem hopeless to compete with them, so that any study at hadron collider
would be doomed to be the second class physics.

Nevertheless, the $B$-physics research at hadron collider does offer several unique possibilities,
which are not available anywhere else. The cross section of $b \bar b$ production at the
Tevatron is significantly higher than that at the $e^+ e^-$ colliders. Therefore, an enormous
statistics of events containing $B$ hadrons is accessible. Of course, not all $B$ events
were recorded by the experiments, but the large number of produced $B$ hadrons allowed to
build highly selective triggers on interesting events and study very rare decays of $B$ hadrons,
such as $\Bd, \Bs \to \mu^+ \mu^-$, or statistically limited processes, like the dimuon charge asymmetry.
The second important feature is the production of all species of $B$ hadrons in $p \bar p$ collisions,
while only $B^\pm$ and $\Bd$ mesons can be studied at the $B$ factories where $B$ hadrons are
mainly created in the decays
of $\Upsilon(4S)$ meson. This possibility allowed the experiments at the Tevatron to perform many
unique measurements, such as the oscillation of $\Bs$ meson, the lifetime and the mass of $\Bcpm$ mesons,
the discovery of baryons containing $b$ quark. The third advantage of the Tevatron
experiments is a large momentum boost of $B$ hadrons. This boost considerably increases their mean
decay length in the laboratory frame.
At $B$ factories, on the contrary, $B$ hadrons are produced at the mass threshold
and their momentum in the laboratory frame is relatively small.
Due to this boost, a more precise measurement of the
$B$-hadron proper lifetime is possible at the Tevatron.
This advantage becomes essential, e.g., for the study of the time evolution of the $\Bs$ system.

It is quite difficult to exploit these advantages to a full extent. The cross section of background
processes in $p \bar p$ collisions at $\sqrt{s} = 1.96$ TeV is $\sim 800$ times larger than
that of the $b \bar b$ production.
The only possibility to suppress this background is to analyse a limited number of
special decay modes of $B$ hadrons, like the decays containing $J/\psi \to \mu^+ \mu^-$ meson,
or simple hadronic decay modes, such as $B_s \to D_s \pi$.
In each event many background particles from the $p \bar p$ interactions accompany
the $B$-hadron decay products. In addition, the Tevatron experiments in their
quest for new physics ran at the highest possible accelerator luminosity, so that
each recorded event, beside the $b \bar b$ production,
contains in average three background pile-up interactions. As a result, an event at hadron
collider looks very complicated. The mean
multiplicity of reconstructed charged particles in a typical recorded event is about 80,
with a very long tail extending above 300 tracks.
Therefore, selecting the charged $B$ hadron decay products among all background tracks results
in a large combinatorial background which should be suppressed by additional requirements.
And due to the presence of a large number of background particles it is almost impossible
to select neutral particles from $B$-hadron decay,\footnote{A rare exception
is the reconstruction of $K_S$ meson and $\Lambda$ baryon using their decays
$K_S \to \pi^+ \pi^-$ and $\Lambda \to p \pi^-$, and of the photon using its conversion
$\gamma \to e^+ e^-$ in the detector material.} since there is no any handle
to associate correctly the clusters in the calorimeters with decaying $B$ hadron.

In addition to these difficulties, the Tevatron detectors were not especially built for
the $B$-physics study, and, e.g., their particle identification was insufficient for many tasks.
The $B$-physics research program competed with other high profile tasks, like the search for the
Higgs boson or the study of top quark. Therefore, the material and human resources available for
$B$-physics studies were limited and these studies were performed by a relatively small number of
dedicated scientists.

All these problems are very serious and disadvantageous, and
nevertheless the $B$-physics research at the Tevatron was proven to be very successful.
One of the important accomplishments of the CDF and D\O\ experiments was namely
this convincing demonstration that,
in spite of all obstacles, performing the $B$-physics research at hadronic
collider is a very rewarding experience producing
unique, world best and valuable results. This success was achieved, among other means,
by a clear definition of research goals, by fully exploiting the advantages provided
by the hadron colliders and detectors, and by pursuing a relatively small number
of high profile measurements which can compete with, or simply cannot be performed at
the $B$ factories.

The aim of this review is to summarize the most important, from the author's point
of view, $B$-physics results obtained by the CDF and D\O\ collaborations.
The Tevatron achievements in $B$ physics provide an excellent
starting point for the LHC experiments at CERN, which continue and extend many studies
started at Fermilab, and which need to overcome similar or even greater problems while
performing the $B$-physics studies. Therefore, the experience learned at the Tevatron
can be useful for the next generation of experiments at hadron colliders, and one of the aims
of this review is to share it. Due to the size limitation, this review is not expected to be
complete and exhaustive, and many excellent results from the Tevatron will stay beyond
the scope of this paper.

After a brief discussion of the CDF and D\O\ detectors in Section \ref{detectors},
with the emphasis on the parts and features essential for the $B$-physics studies, I will review
the discovery of the new $B$ hadrons in Section \ref{discovery} and
the measurement of the $B$-hadron lifetimes in Section \ref{lifetimes}. The breakthrough
measurement of the oscillation frequency of the $\Bs$ meson is presented
in Section \ref{mixing}. The study of the decays of $B$ mesons, including the search for the rare
decays providing strong constraint on the new physics contribution is given in Section \ref{rare}.
Finally, the study of the $CP$ asymmetry in $B$-meson decays and mixing is discussed in Section \ref{cpv}.

\section{Detectors}
\label{detectors}

The CDF and D\O\ experiments are the general purpose collider detectors constructed to maximally exploit the possibilities
provided by the $p \bar p$ collisions at $\sqrt{s} = 1.96$ TeV and to operate at the instantaneous luminosity up to
$5 \times 10^{32}$ cm$^{-2}$ s$^{-1}$. Although the main emphasis in their design is made on the detection
of events with the highest possible invariant mass, they also
contain the specific elements necessary to endeavour the
$B$-physics research.
%Both these detector have the central cylindrical structure and two end-caps,
%which cover almost the full solid angle.

The tracking system of the CDF detector\cite{detector-cdf} includes the solenoidal magnet producing a uniform magnetic
field of 1.4 T, the inner tracking volume containing the silicon microstrip detectors up to a radius
of 28 cm from the beamline\cite{silicon-cdf}, and the outer tracking volume
instrumented with an open-cell drift chamber\cite{tracker-cdf} (COT) up to the radius of 137 cm.
The first single-sided layer of the silicon detector\cite{l0-cdf} is mounted directly on the beam-pipe at the
radius of 1.5 cm. The tracking systems reconstructs the trajectory and momentum of the charged particles up to the
pseudorapidity\footnote{ The pseudorapidity is defined as $\eta = -\ln(\tan(\theta / 2))$, where $\theta$ is the
polar angle of charged track relative to the direction of the proton beam.}
$|\eta| < 2.$
The resolution of the track impact parameter\footnote{The impact parameter is defined as the distance
of closest approach of the charged particle to the $p \bar p$ interaction point in the plane
transverse to the beam direction.} is about 40 $\mu$m. This resolution includes an
uncertainty of the interaction point in the transverse plane, which is about 30 $\mu$m.
The momentum resolution of the tracking system is $\sigma(p_T)/p_T^2 \simeq 1.7 \times 10^{-3}$ GeV$^{-1}$,
where $p_T$ is the component of the particle momentum transversal to the beam direction.

The muon identification system\cite{muon-cdf,muon-cdf-1} is located after the magnet and the calorimeters, which serve as a shield
to suppress the penetration of all charged and neutral particles except the muons. It includes
the drift chambers, which detect muons with $p_T > 1.4$ GeV within $|\eta| < 0.6$,
and additional chambers and scintillators,
which cover $0.6 < |\eta| < 1.0$ for muons with $p_T > 2.0$ GeV.

An important component of the CDF detector essential for the $B$-physics measurements is the special
trigger selecting events with displaced tracks. It is a three-level system. At the first level\cite{level1-cdf}
the COT hits are grouped into tracks in the transverse plane.
At the second level\cite{level2-cdf}, the silicon hits are added to the tracks found at the first level.
These hits improve the resolution of the track impact parameter, which is measured in real time.
Finally, the displaced vertex trigger\cite{level3-cdf} requires two charged particles
with $p_T > 2$ GeV, and with impact parameters in the range $0.12 - 1$ mm.
This trigger configuration is the basis for many CDF
measurements with fully hadronic $B$ decays. Its other trigger configurations select
the events with one or two muons.
%The strong side of the D\O\ detector is its muon identification system, which allows an
%independent measurement of the muon momentum and provides a very clean muon selection with strong
%background suppression. However, its trigger system does not allow
%to select events with the displaced tracks.
%

The central tracking system of the D\O\ detector\cite{detector-d0}
comprises a silicon microstrip tracker and a central fiber tracker, both located within a 1.9 T superconducting
solenoidal magnet. The outer radius of the tracking system is 52 cm. The tracking system reconstructs
tracks with $|\eta| < 2.2$. The closest to the beam layer of the silicon detector\cite{l0-d0,l0-d0-1} is
located at the radius 1.7 cm. The impact parameter precision of high momentum tracks is $\sim 18$ $\mu$m.
The momentum resolution provided by the tracking system is $\sigma(p_T)/p_T^2 \simeq 3.0 \times 10^{-3}$ GeV$^{-1}$
for tracks with $p_T > 5$ GeV.

The muon system\cite{muon-d0} is located beyond the calorimeters that surround the central
tracking system, and consists of a layer of tracking detectors
and scintillation trigger counters before 1.8 T iron
toroids, followed by two similar layers after the toroids. It is able to identify the muons with $|\eta| < 2.0$.
The toroidal magnets allow an independent measurement of the muon momentum, which
helps to improve the quality of identified muons.

The trigger system of the D\O\ detector does not provide a possibility to collect events
with displaced tracks, although its muon and di-muon triggers
are very efficient and robust.
Therefore the focus
of the $B$-physics measurements in D\O\ experiment is shifted towards the
semileptonic $B$ decays and decays with $J/\psi \to \mu^+ \mu^-$ in the final state.

The polarities of the toroidal and solenoidal magnetic fields of the D\O\ detector are regularly reversed.
This reversal helps to significantly reduce the systematic uncertainties of the measurements sensitive
to the differences
in the reconstruction efficiency between the positive and negative particles, like the measurements of the
$CP$ violating charge asymmetries.

Thus, both experiments have sufficient
and powerful  tools to fulfil their $B$-physics research program.
They also contain several special features
which make them different and complementary.  The CDF detector has
a larger tracking volume. Therefore its charged particle momentum resolution is superior to that of the
D\O\ detector. It also can select the hadronic $B$ decays.
The D\O\ detector includes a sophisticated muon identification system
with local measurement of muon momentum. It extends up to $|\eta < 2.0|$ and provides
a clean selection of muons with strong background suppression. The reversal of magnet polarities
allows it to perform several measurements of the charge asymmetry in the semileptonic $B$ decays which are
at the world best level.

\section{Discovery of New Particles containing $b$ quark}
\label{discovery}

The search for new particles is always a very exciting task. Historically, namely such a search
helped to establish, e.g., the quark model or the Standard Model (SM) of
electro-weak interactions as true theories. The quark model describes the observed
hadrons as the bound states of quarks and anti-quarks. It predicts in particular the spectrum of particles
containing $b$ quark. Various theoretical techniques based on the quark model
are developed to compute the properties of these particles.
Comparison of these predictions with the experimental observations helps
to improve and rectify these techniques and develop new more powerful methods. Therefore, discovery
of new particles containing $b$ quarks constitutes an important part of $B$ physics.
An observation
of each new object attracts a lot of attention
in particle physics community, and each such observation provides
a new confirmation of the validity and strength of the quark model.
Since all types of $B$ hadrons can be produced at hadron
colliders with relatively high production rate, it is an ideal place for this kind of research.
The discovery of several such objects by the experiments at the Tevatron
is described in this section.

\subsection{Mass of $\Bs$ meson and $\Lb$ baryon}

Although the $\Bs$ meson and $\Lb$ baryon were first observed at LEP \cite{pdg-2012}, their
mass was first precisely measured
by the CDF collaboration. The CDF measured simultaneously the masses
of $\Bp$, $\Bd$, $\Bs$ mesons and $\Lb$ baryon \cite{bmass-cdf}. The comparison of the $\Bp$ and $\Bd$ mass
with the results obtained at the $e^+e^-$ collider by the CLEO collaboration \cite{bmass-cleo}
provides an excellent cross-check
of the measurement technique and gives a confidence in the values of the $\Bs$ and $\Lb$ mass, which was
known with a very poor precision at that time. The obtained results are:
\begin{eqnarray}
m(\Bp) & = & 5279.10 \pm 0.41\mbox{(stat)} \pm 0.36\mbox{(syst)}~ {\rm MeV}, \\
m(\Bd) & = & 5279.63 \pm 0.53\mbox{(stat)} \pm 0.33\mbox{(syst)}~ {\rm MeV}, \\
m(\Bs) & = & 5366.01 \pm 0.73\mbox{(stat)} \pm 0.33\mbox{(syst)}~ {\rm MeV}, \\
m(\Lb) & = & 5619.7~ \pm 1.2~\mbox{(stat)} \pm 1.2~\mbox{(syst)}~ {\rm MeV}.
\end{eqnarray}
The achieved precision of the $\Bp$ and $\Bd$ mass is compatible or even better than the
CLEO values, while the precision of the $\Bs$ and $\Lb$ mass is improved by more than 10 times
compared to the previous LEP measurements. Recently, the LHCb collaboration obtained the most precise
values of the $B$-hadron masses \cite{bmass-lhcb}, which agree well with the CDF results.

\subsection{$\Bcpm$ meson}

The $\Bcm$ mesons contains $b$ quark and $c$ anti-quark. A relatively heavy mass
of these quarks simplifies the theoretical calculations of the $\Bcpm$ properties.
For lattice QCD model, the $\Bcpm$ is called a ``gold-plated" hadron \cite{Brambilla}.
However, various theoretical models predict different mass ranges for this particle
with large uncertainties. Non-relativistic quark models \cite{Quigg,Quigg-1,Quigg-2} predict the $\Bcpm$ mass
in the range 6247 -- 6286 MeV. The lattice QCD calculations\cite{Allison} provide
a slightly different mass value $M(\Bcpm) = 6304 \pm 12 ^{+18}_{-0}$ MeV. Therefore,
a precise measurement of the $\Bcpm$ mass is essential for verifying these calculations
and improving the computational techniques.

\begin{figure}[tpbh]
\begin{center}
%\centerline{
\hspace*{-0.15\textwidth}
\epsfig{figure=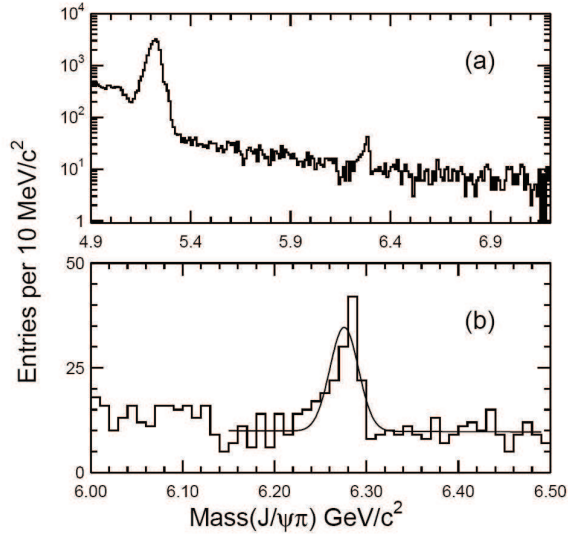,width=0.5\textwidth}
%}
%\includegraphics[width=0.80\textwidth]{bc-mass-cdf.eps}
\vspace*{8pt}
\caption{
(a). The invariant mass distribution of $J/\psi K^\pm$ combinations from Ref. \cite{bc-mass-cdf}.
(b) Identical to (a), but in a narrower mass range.
The projection of
the fit to the data is indicated by the curve overlaid on (b).
\label{fit-bc-cdf}}
\end{center}
\end{figure}

The first precise measurement of the $\Bcpm$ mass was done in the decay mode $\Bcpm \to J/\psi \pi^\pm$,
with $J/\psi \to \mu^+ \mu^-$, by the CDF collaboration \cite{bc-mass-cdf} using the data
corresponding to an integrated luminosity of 2.4 fb$^{-1}$ .
It is the easiest decay mode to separate the $\Bcpm$ production from background because
of the clean muon identification and the pure $J/\psi \to \mu^+ \mu^-$ decay selection.
A set of cuts to select the signal was developed and tested using the decay $\Bpm \to J/\psi K^\pm$,
which has a similar final state topology with the only difference that the $\pi^\pm$ meson
in $\Bcpm$ decay is replaced by the $K^\pm$ meson. Since the lifetime of $\Bcpm$ meson
is much shorter than the lifetime of $\Bpm$ meson, only the $\Bpm \to J/\psi K^\pm$ candidates
with proper decay length $80 < ct < 300$ $\mu$m were used to test the selection criteria.
The mass distribution of the selected $\Bcpm \to J/\psi \pi^\pm$ candidates is shown in Fig.\ \ref{fit-bc-cdf}.
An excess of $J/\psi \pi^\pm$ events with the invariant mass around 6280 MeV is clearly seen.
The unbinned fit over the mass range 6150 −- 6500 MeV gives a $\Bcpm$ signal of $108 \pm 15$
candidates with a mass of
\begin{equation}
m(\Bcpm) = 6275.6 \pm 2.9\mbox{(stat)} \pm 2.5\mbox{(syst)}~ {\rm MeV}.
\end{equation}
The statistical significance of this observation corresponds to 8 standard deviations.

\begin{figure}[tpbh]
\begin{center}
\centerline{
\psfig{file=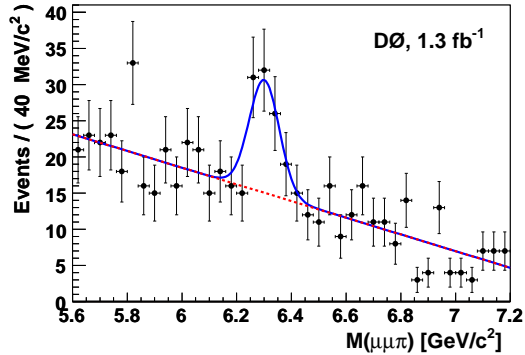,width=0.6\textwidth}
}
\vspace*{8pt}
\caption{
The $J/\psi \pi^\pm$ invariant mass distribution of $\Bcpm$ candidates from Ref. \cite{bc-mass-d0}.
A projection of the unbinned maximum
likelihood fit to the distribution is shown overlaid.
\label{fit-bc-d0}}
\end{center}
\end{figure}

The D\O\ collaboration performed a similar analysis of the $J/\psi \pi^\pm$ final state \cite{bc-mass-d0}.
They used the integrated luminosity 1.3$^{-1}$ fb. The resulting invariant mass distribution of
the selected $J/\psi \pi^\pm$ combinations is shown in Fig.\ \ref{fit-bc-d0}. The unbinned fit of
selected events gives $54 \pm 12$ events and the $\Bcpm$ mass
\begin{equation}
m(\Bcpm) = 6300 \pm 14\mbox{(stat)} \pm 5\mbox{(syst)} ~{\rm MeV}.
\end{equation}
The statistical significance of the observed $\Bcpm$ signal
exceeds 5 standard deviations above the background level.

Comparing the results of the two experiments, it is clearly seen that the mass resolution in
the CDF experiments is much better, which is reflected in a smaller uncertainty of the measured
$\Bcpm$ mass and smaller background level for the CDF measurement.
This difference is mainly caused by $\sim 2.5$ times larger tracking volume of the CDF detector,
which gives a significant gain in the momentum resolution of the charged particles.
%which more than compensate a smaller magnetic field.
Nevertheless, the result of D\O\ experiment
provides an important confirmation of the CDF observation. The two measurements of the $\Bcpm$
mass agree within 2 standard deviations, so that the existence of the $\Bcpm$ is firmly established.
The recent result of the LHCb experiment \cite{bc-mass-lhcb}, which obtains
%the $\Bcpm$ mass
\begin{equation}
m(\Bcpm) = 6273.7 \pm 1.3\mbox{(stat)} \pm 1.6\mbox{(syst)} {\rm MeV},
\end{equation}
confirms these two observations.

\subsection{$\Sigma_b^\pm$ and $\Sigma_b^{*\pm}$ baryons}

The quark model predicts the existence of many baryons containing
$b$ quark,
%\cite{Jenkins,Mathur,Ebert,Karliner}
but only the $\Lambda_b$
baryon with quark content $(bdu)$ was observed before the start of the experiments at the Tevatron.
The search for other $B$ baryons at CDF and D\O\ experiments was very fruitful, as essentially all
baryons with one $b$ quark, except $\Sigma^0_b$ baryon with quark content (bdu), were
discovered by them. The theory also predicts the mass of all these objects and other
properties which can be tested experimentally, although different models
give slightly different values. Therefore, an observation of these objects
provides direct comparison with theory, which is essential for verifying
theoretical models and helping to improve them.

The quark content of $\Sigma_b^{(*)+}$
baryon\footnote{The notation $\Sigma_b^{(*)\pm}$ refers to the states
$\Sigma_b^{\pm}$ and $\Sigma_b^{*\pm}$.}
is $(buu)$ and the quark content of
$\Sigma_b^{(*)-}$ baryon is $(bdd)$. Thus, they are different particles with different mass.
In the heavy quark effective
theory (HQET)\cite{hqet}, describing these objects,
a heavy $b$ quark is considered as static and surrounded
by a diquark system of two light quarks. This diquark system has isospin $I = 1$
and $J^P = 1^+$, so that the lightest quark systems $(buu)$ and $(bdd)$ can have the
quantum numbers $J^P = \frac{1}{2}^+$ ($\Sigma_b^\pm$) and
$J^P = \frac{3}{2}^+$ ($\Sigma_b^{*\pm}$). An extensive list of theoretical papers,
predicting the masses of $\Sigma_b^{(*)\pm}$ baryons
can be found in Ref. \cite{sigmab-2}, here we give just some of
them.\cite{Jenkins,Mathur,Mathur-1,Ebert,Ebert-1,Ebert-2,Karliner,Rosner} Based on these works, the mass difference
between $\Sigma_b^{(*)\pm}$ and $\Lambda_b$ baryons is expected to be
$m(\Sigma_b) - m(\Lambda_b) \sim 180 - 210$ MeV; therefore, the main
decay mode of  $\Sigma_b^{(*)\pm}$ should be $\Sigma_b^{(*)\pm} \to \Lambda_b \pi^\pm$.

The observation of these particles were first reported by the CDF collaboration in Ref. \cite{sigmab-1},
and the following paper\cite{sigmab-2} gives an improved measurement of their mass using
the integrated luminosity 6 fb$^{-1}$. The $\Sigma_b^{(*)\pm}$ baryons were searched for in the decay
$\Sigma_b^{(*)\pm} \to \Lambda_b \pi^\pm$, with $\Lambda_b \to \Lambda_c^+ \pi^-$ and
$\Lambda_c^+ \to p K^- \pi^+$. The decays were selected with the displaced
two-track trigger, which allowed the CDF experiment to collect the events with hadronic
decays of heavy hadrons. There was not such a possibility in the D\O\ detector and
it does not have any result on the studies involving the hadronic decays of $B$ hadrons.

\begin{figure}[tpbh]
\begin{center}
\centerline{
\psfig{file=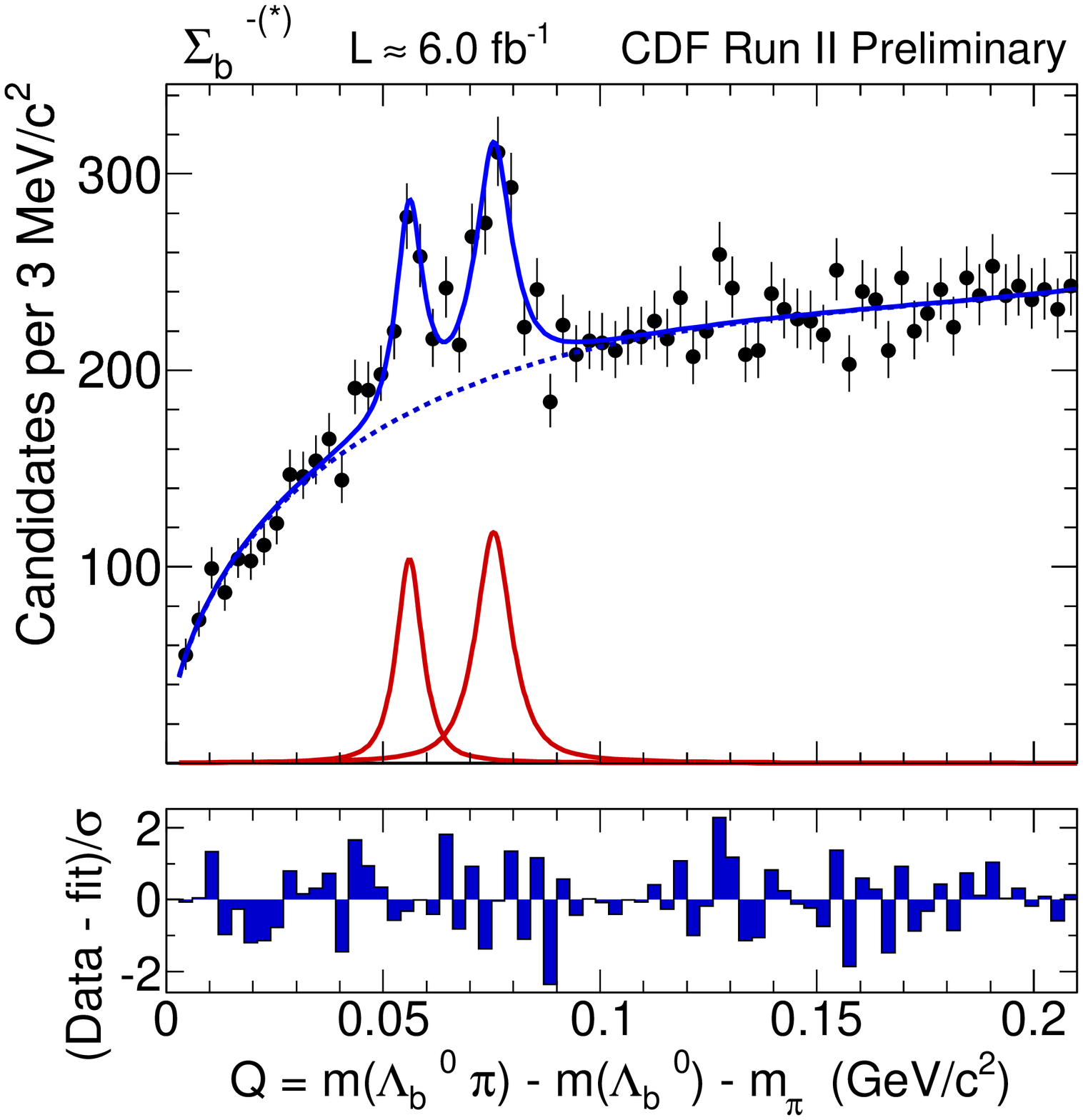,width=0.50\textwidth}
\psfig{file=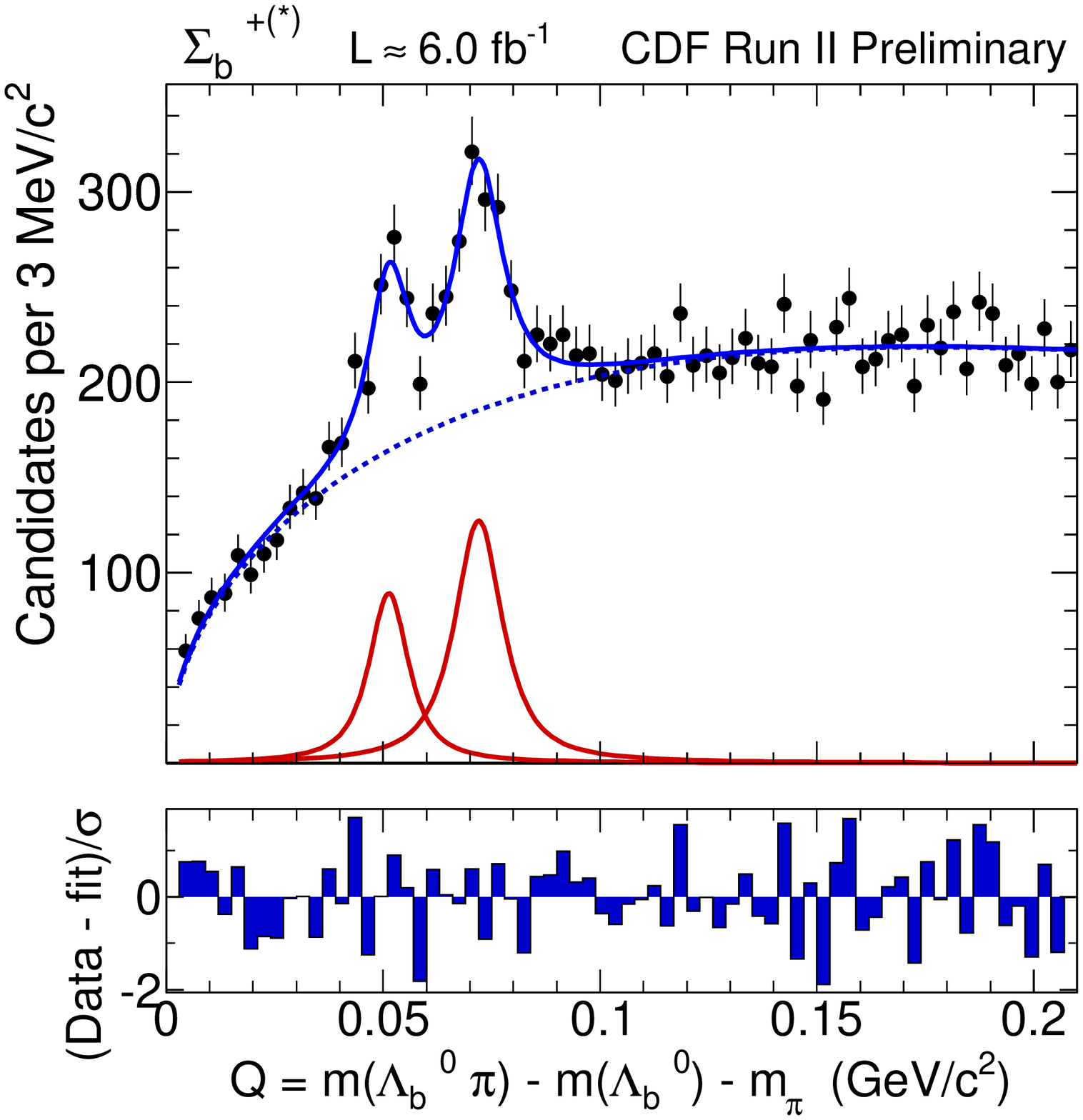,width=0.50\textwidth}
}
\vspace*{8pt}
\caption{
The left (right) plot shows the $Q$-value spectrum for $\Sigma_b^-$ ($\Sigma_b^+$)
candidates from Ref. \cite{sigmab-2} with the projection of the corresponding
unbinned likelihood fit superimposed.
The $Q$ value is defined in Eq. (\ref{qvalue}).
The pull distribution of each fit is shown in the bottom of the corresponding plot.
\label{sigmab}}
\end{center}
\end{figure}

Using the sample of approximately 16300 reconstructed $\Lambda_b$ decays, the CDF collaboration
built the mass difference
\begin{equation}
Q = m(\Lambda_b \pi) - m(\lambda_b) - m(\pi),
\label{qvalue}
\end{equation}
which is shown in
Fig.\ \ref{sigmab}. The signals of both $\Sigma_b^\pm$ and $\Sigma_b^{*\pm}$ are clearly seen.
The parameters of all four baryons obtained from the fit of these distributions are given
in Table \ref{tab1}.

\begin{table}[tbph]
\begin{center}
\caption{Parameters of $\Sigma_b^{(*)\pm}$ baryons from Ref. \cite{sigmab-2}.
The first uncertainty is statistical and the second is systematic.
}
{\begin{tabular}{@{}|c|c|c|c|@{}} 
\hline
State & $Q$ value (MeV) & Mass $m$ (MeV) & Natural width $\Gamma$ (MeV) \\
\hline
$\Sigma_b^{-\hphantom{*}}$ & $56.2^{~+0.6~+0.1}_{~-0.5 ~ -0.4}$ &
$5815.5^{+0.6}_{-0.5} \pm 1.7$  & $\hphantom{0}4.9^{+3.1}_{-2.1} \pm 1.1$ \\
                           &                                    &                    &      \\
$\Sigma_b^{*-}$            & $75.8 \pm 0.6 ^{+0.1}_{-0.6}$      &
$5835.1 \pm 0.6 ^{+1.7}_{-1.8}$ & $\hphantom{0}7.5^{~+2.2 ~+0.9}_{~-1.8 ~-1.4}$ \\
                           &                                    &                    &      \\
$\Sigma_b^{+\hphantom{*}}$ & $52.1^{~+0.9~+0.1}_{~-0.8 ~ -0.4}$ &
$5811.3^{+0.9}_{-0.8} \pm 1.7$  & $\hphantom{0}9.7^{~+3.8 ~+1.2}_{~-2.8 ~-1.1}$ \\
                           &                                    &                    &      \\
$\Sigma_b^{*+}$            & $72.8 \pm 0.7 ^{+0.1}_{-0.6}$      &
$5832.1 \pm 0.7 ^{+1.7}_{-1.8}$ & $11.5^{~+2.7 ~+1.0}_{~-2.2 ~-1.5}$ \\ 
\hline
\end{tabular}
\label{tab1}}
\end{center}
\end{table}

In addition, the CDF collaboration measured the isospin mass splitting between the positive and negative
$\Sigma_b$ baryons:
\begin{eqnarray}
m(\Sigma_b^+) - m(\Sigma_b^-) & = & −4.2^{+1.1}_{-1.0} \pm 0.1 ~\mbox{MeV}, \\
m(\Sigma_b^{*+}) - m(\Sigma_b^{*-}) & = & −3.0^{+1.0}_{-0.9} \pm 0.1 ~\mbox{MeV}.
\end{eqnarray}
Thus, the $\Sigma_b^{(*)-}$ baryons are heavier than the $\Sigma_b^{(*)+}$ baryons,
which is similar to the pattern observed in all other isospin multiplets. It can be
explained by the larger mass of the $d$ quark with respect
to the $u$ quark.  The electromagnetic contribution due to electrostatic Coulomb forces between quarks,
which is larger in $\Sigma_b^{(*)-}$ baryons than in $\Sigma_b^{(*)+}$ baryons, can also contribute
in the observed mass splitting.

\subsection{$\Xibpm$ baryon}

\begin{figure}[tpbh]
\begin{center}
\centerline{
\psfig{file=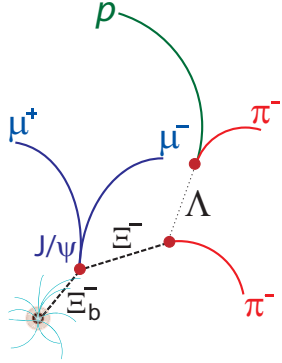,width=0.30\textwidth}
}
\vspace*{20pt}
\caption{
Schematic of the $\Xibm \to J/\psi \Xi^- \to J/\psi \Lambda \pi^- \to
(\mu^+ \mu^-) (p \pi^-) \pi^-$ decay topology taken from Ref. \cite{xib-d0}. The $\Lambda$
and $\Xi^-$ baryons have decay lengths of the order of cm; the $\Xibm$ baryon
has a decay length of the order of mm.
\label{xib-decay}}
\end{center}
\end{figure}

The $\Xibm$ baryon with the quark content ($bds$) contains the down-type quarks from all three families.
Since it contains the $s$ quark, it decays weakly and has a relatively long
lifetime with its decay vertex separated from the primary interaction. It was first discovered
by D\O\ collaboration\cite{xib-d0}  and confirmed by the CDF collaboration\cite{xib-cdf} soon
after that. It was observed in the decay mode
$\Xibm \to J/\psi \Xi^-$ with $J/\psi \to \mu^+ \mu^-$, $\Xi^- \to \Lambda \pi^-$, and $\Lambda \to p \pi^-$.
Reconstruction of this decay chain is quite difficult because it contains three separate
decay vertices as can be seen in Fig.\ \ref{xib-decay}.

\begin{figure}[tpbh]
\begin{center}
\centerline{
\psfig{file=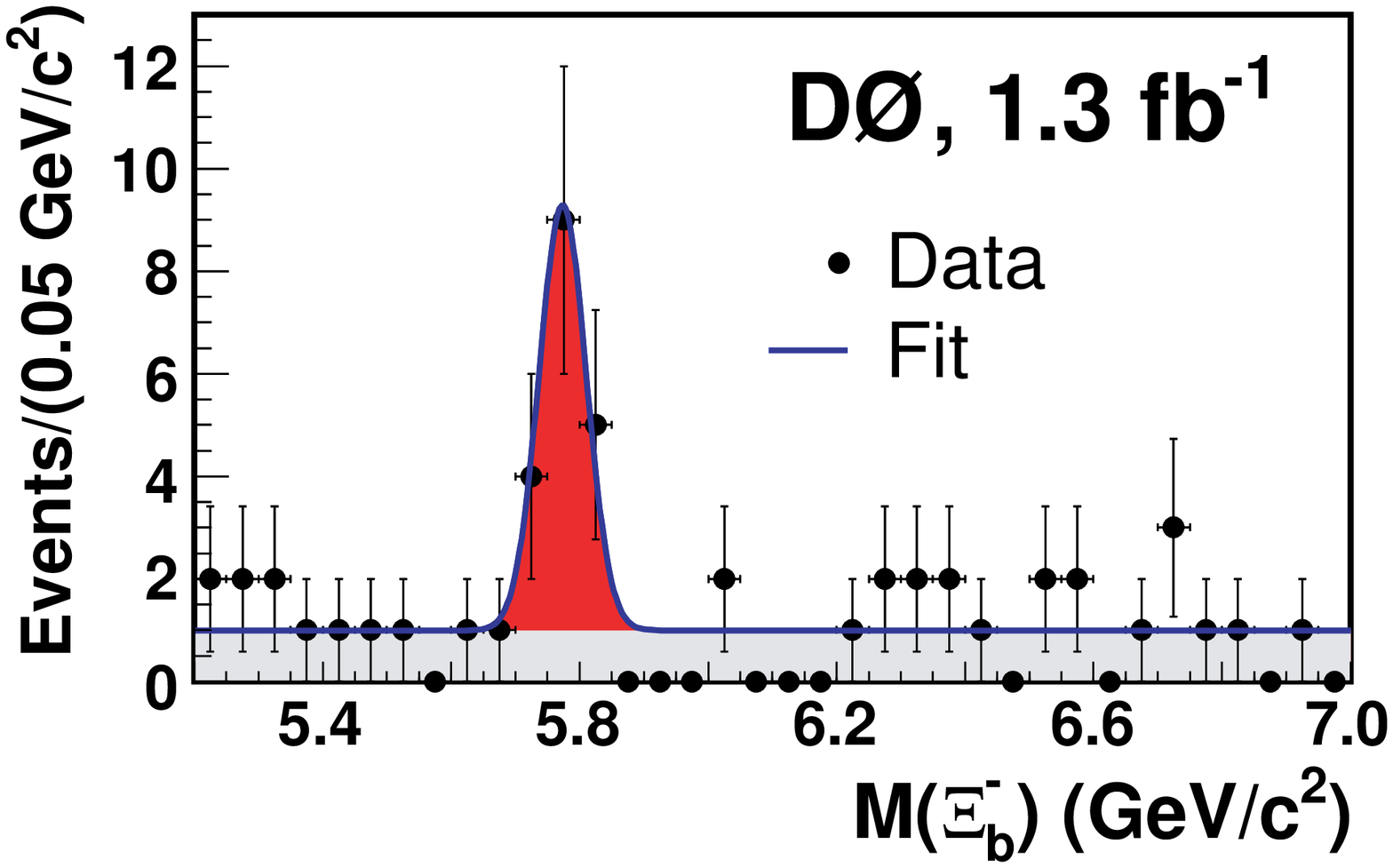,width=0.50\textwidth}
\psfig{file=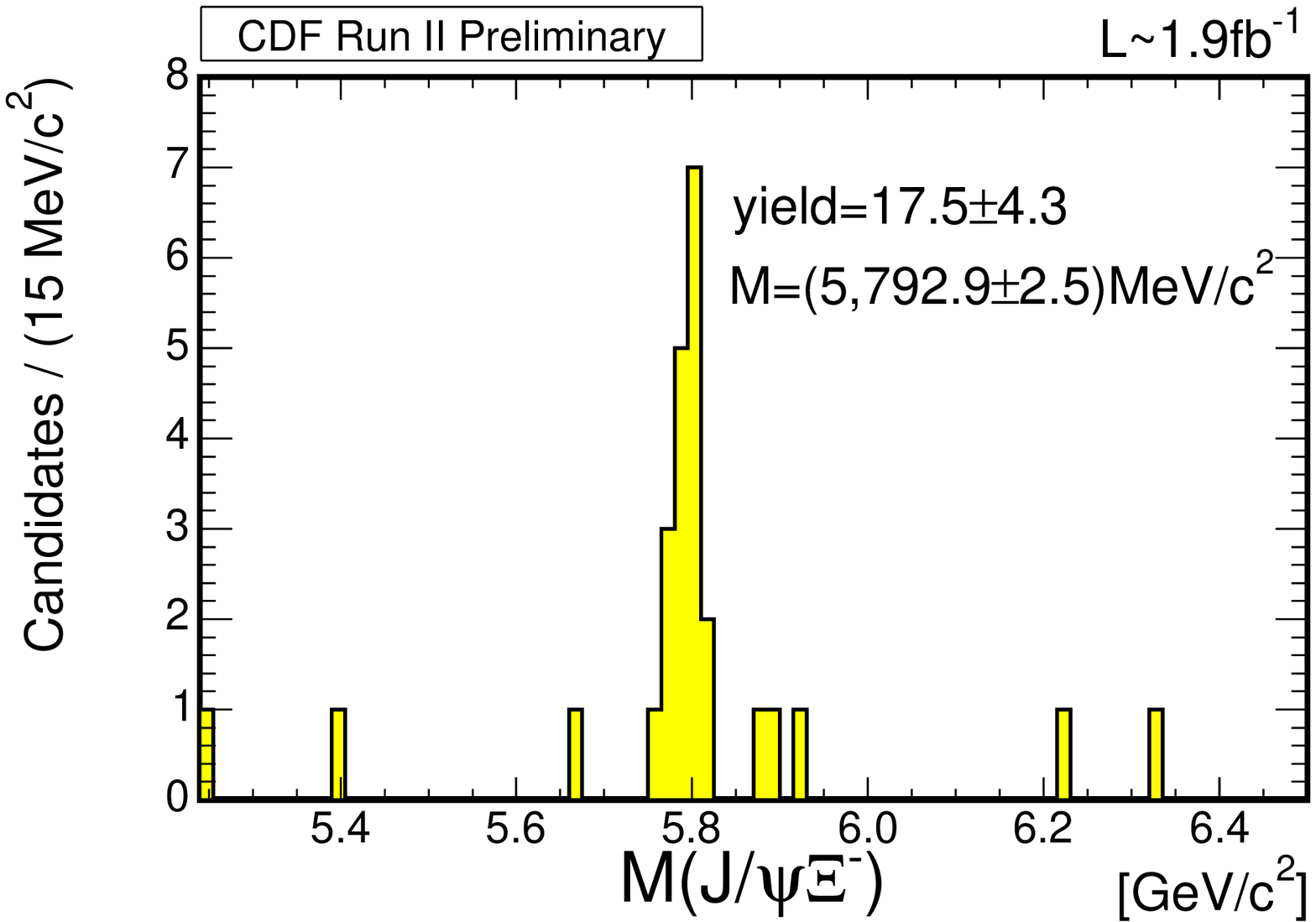,width=0.50\textwidth}
}
\vspace*{8pt}
\caption{
The left (right) plot shows the $J/\psi \Xi^\pm$ invariant mass distribution
obtained by the D\O\ (CDF) collaboration, and taken from Ref. \cite{xib-d0} (\cite{xib-cdf}), respectively.
\label{xib-mass}}
\end{center}
\end{figure}

Both collaborations developed sophisticated selections to suppress a background and obtain
a clean signal. In particular, they utilize the relatively long lifetime of $\Xibpm$ baryon
and a distinctive separation of its decay vertex from the primary interaction point. As a result,
the suppression of background for this final state is strong, and a clean $\Xibpm$ signal is
obtained. The left plot in Fig.\ \ref{xib-mass} shows the $J/\psi \Xi^\pm$ mass distribution obtained by the
D\O\ collaboration after all selection criteria applied, while the right plot shows the similar
mass distribution obtained by the CDF collaboration. In both cases, a clean and statistically significant
signal of the $\Xibpm$ production is observed.

The measured mass of the $\Xibpm$ baryon is found to be
\begin{eqnarray}
m(\Xibpm) & = & 5774 \pm 11\mbox{(stat)} \pm 17\mbox{(syst)} ~\mbox{MeV (D\O\ Collab.)}, \\
m(\Xibpm) & = & 5792.9 \pm 2.5\mbox{(stat)} \pm 1.7\mbox{(syst)} ~\mbox{MeV (CDF Collab.)}.
\end{eqnarray}
Again, a better precision of the CDF measurement is mainly due to a better precision of the tracking
system of the CDF detector. In the most recent measurement\cite{omb-cdf} the CDF collaboration
reports a statistically consistent value of the mass with the improved systematic uncertainty
\begin{equation}
m(\Xibpm)  =  5790.9 \pm 2.6\mbox{(stat)} \pm 0.8\mbox{(syst)} ~\mbox{MeV}.
\end{equation}
The obtained $\Xibpm$ mass is in a qualitatively good agreement with the
theoretical predictions\cite{Jenkins,Mathur,Mathur-1,Ebert,Ebert-1,Ebert-2,Karliner}.

\subsection{$\Xibn$ baryon}

The $\Xibn$ baryon is the isospin partner of the $\Xibm$ baryon and has the quark content $(usb)$.
Contrary to $\Xibm$ baryon, there is no $\Xibn$ decay mode containing $J/\psi \to \mu^+ \mu^-$
and charged particles, therefore it can not be studied in the D\O\ experiment.
It is observed by the CDF collaboration\cite{xib0-cdf} in the decay mode $\Xibn \to \Xi_c^+ \pi^-$, where
$\Xi_c^+ \to \Xi^- \pi^+ \pi^+$, $\Xi^- \to \Lambda \pi^-$, and $\Lambda \to p \pi^-$.
The data sample used for this observation corresponds to an integrated luminosity of 4.2 fb$^{-1}$
and is collected using the displaced two-track trigger. In parallel, the same analysis
searched for the decay $\Xibm \to \Xi_c^0 \pi^-$, with $\Xi_c^0 \to \Xi^- \pi^+$. It was selected
using similar criteria and the same trigger. Since the $\Xibm$ baryon was established
by the time of this study, this decay provides an excellent cross check of the full
analysis chain applied in the search for the new particle.

\begin{figure}[tpbh]
\begin{center}
\centerline{
\psfig{file=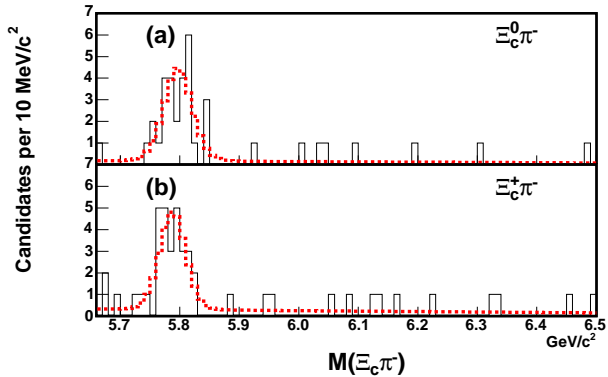,width=0.70\textwidth}
}
\vspace*{8pt}
\caption{
(a) The $\Xibm \to \Xi_c^0 \pi^-$ and (b) the $\Xibn \to \Xi_c^+ \pi^-$ mass distributions
from Ref.\ \cite{xib0-cdf}. A projection of the likelihood fit is overlaid as a dashed line.
\label{xib0-mass}}
\end{center}
\end{figure}

The obtained distributions of the $\Xi_c^+ \pi^-$ and $\Xi_c^+ \pi^- \pi^-$ invariant mass
are shown in Fig.\ \ref{xib0-mass}. The signals of both $\Xibm \to \Xi_c^0 \pi^-$ and
$\Xibn \to \Xi_c^+ \pi^-$ decays are clearly seen. In total, $25.3^{+5.6}_{-5.4}$ candidates
of the $\Xibn \to \Xi_c^+ \pi^-$ decay are observed with the significance greated than
6 standard deviations. The measured mass of $\Xibn$ baryon is found to be
\begin{equation}
m(\Xibn) = 5787.8 \pm 5.0\mbox{(stat)} \pm 1.3\mbox{(syst)} ~\mbox{MeV}.
\end{equation}
In addition, the first observation of the $\Xibm \to \Xi_c^0 \pi^-$ decay is reported
with $25.8^{+5.5}_{-5.2}$ candidates selected.

\subsection{$\Ombpm$ baryon}

The $\Ombm$ baryon contains $(bss)$ quarks. It was observed by the D\O \cite{omb-d0} and
CDF \cite{omb-cdf} collaborations in the decay mode
$\Ombm \to J/\psi \Omega^-$ with $J/\psi \to \mu^+ \mu^-$, $\Omega^- \to \Lambda K^-$,
and $\Lambda \to p \pi^-$. This decay mode is similar to the decay mode used for the
$\Xibpm$ discovery with $\Xi^- \to \Lambda \pi^-$ replaced by the $\Omega^- \to \Lambda K^-$
decay, i.e., the same group of charged particles and the same vertex topology is selected
as for the search for $\Xibpm$, with one of the particles assigned the mass of kaon instead of pion.

\begin{figure}[tpbh]
\begin{center}
\centerline{
\hspace*{0.5cm}\psfig{file=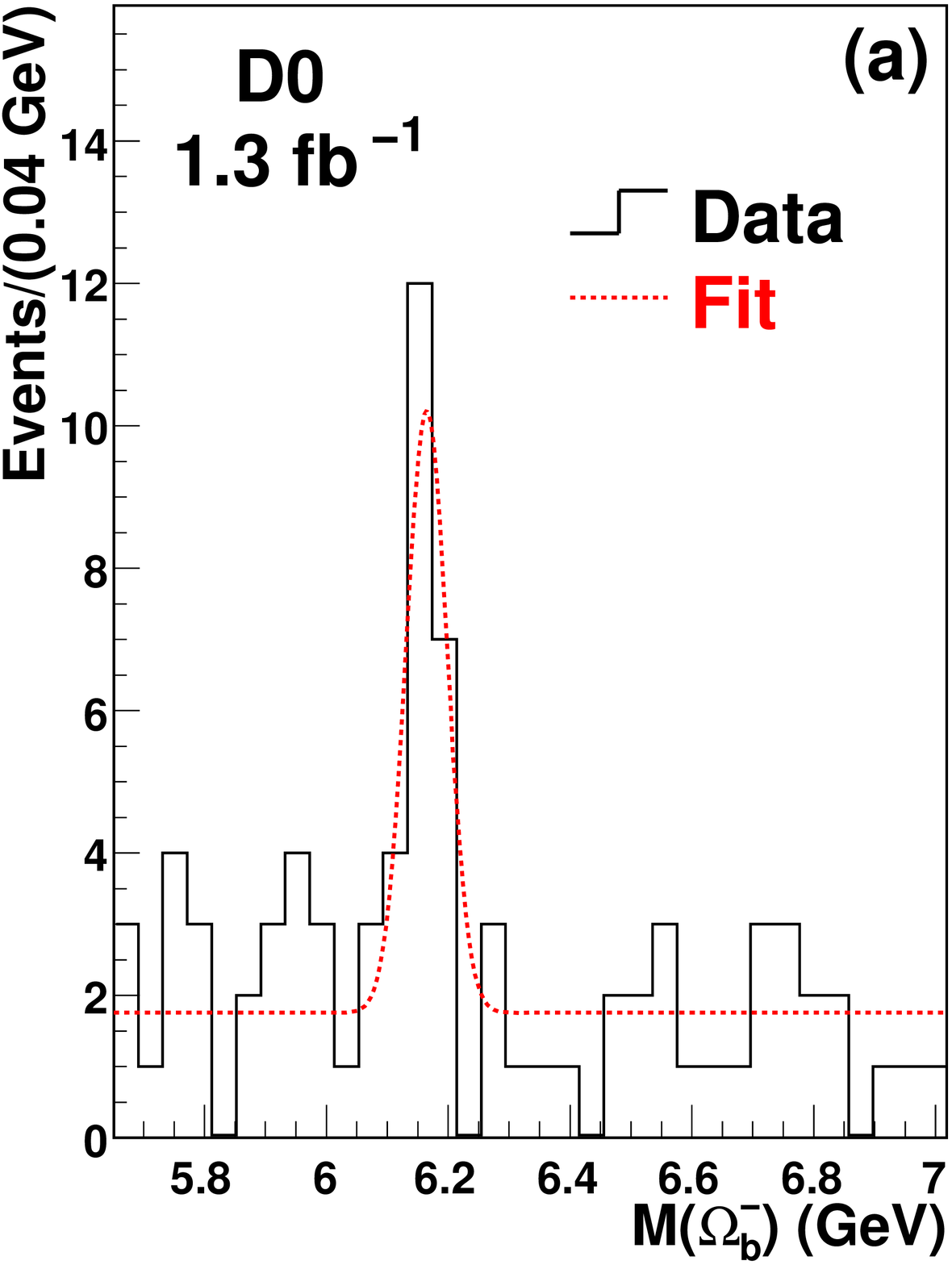,width=0.40\textwidth}
\psfig{file=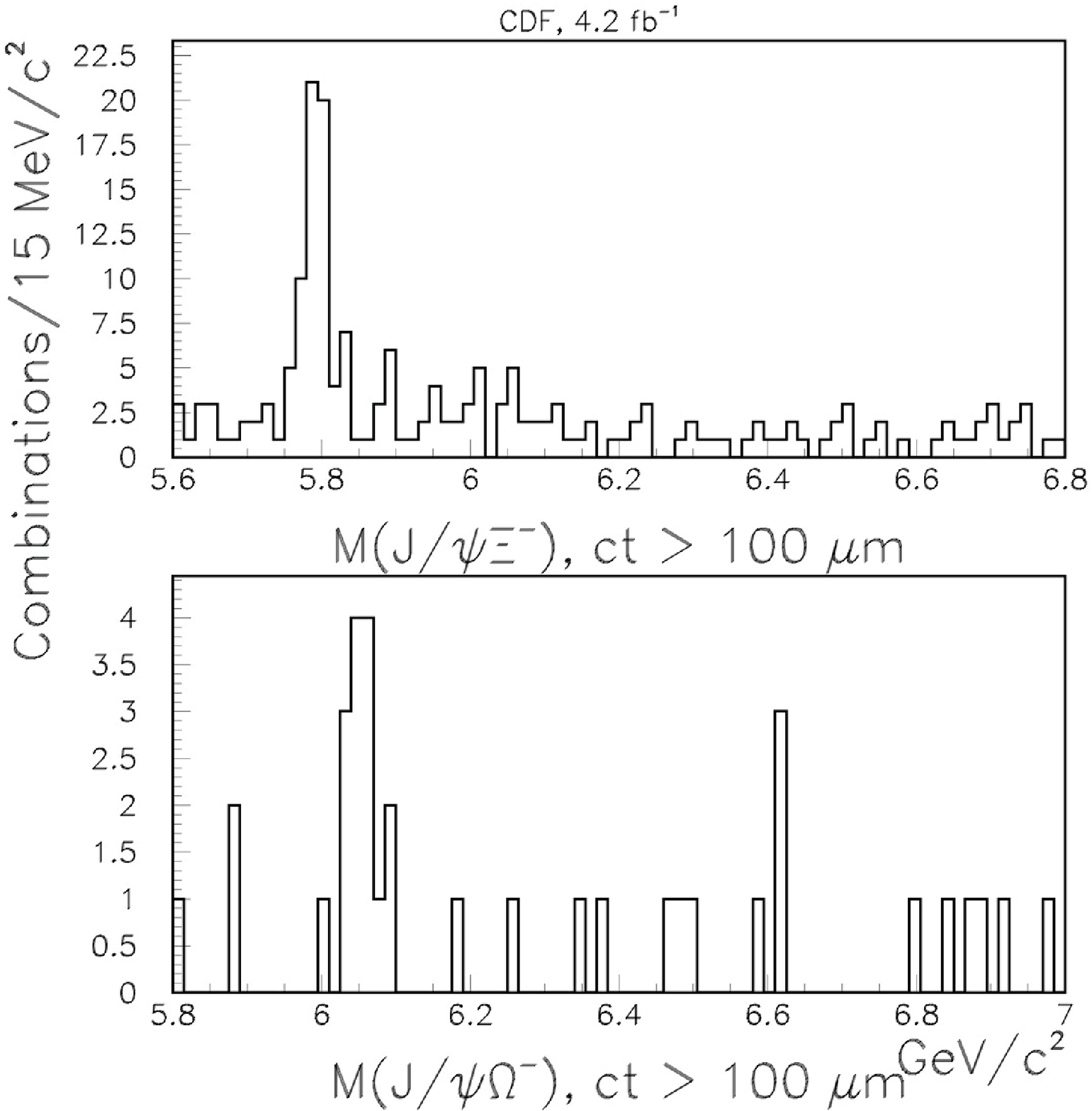,width=0.60\textwidth}
}
\vspace*{8pt}
\caption{
The left (right) plot shows the $J/\psi \Omega^\pm$ invariant mass distribution
obtained by the D\O\ (CDF) collaboration, and taken from Ref. \cite{omb-d0} (\cite{omb-cdf}), respectively.
The control distribution
of the $J/\psi \Xi^\pm$ invariant mass obtained by the CDF collaboration is also shown.
\label{omb-mass}}
\end{center}
\end{figure}

The results of both experiments are shown in Fig.\ \ref{omb-mass}.
The measured mass of the $\Ombpm$ baryon is found to be
\begin{eqnarray}
m(\Ombpm) & = & 6165 \pm 10\mbox{(stat)} \pm 15\mbox{(syst)} ~\mbox{MeV (D\O\ Collab.)}, \\
m(\Ombpm) & = & 6054.4 \pm 6.8\mbox{(stat)} \pm 0.9\mbox{(syst)} ~\mbox{MeV (CDF Collab.)}.
\end{eqnarray}
There is a clear disagreement between these two measurements with the difference between them
$111 \pm 12\mbox{(stat)} \pm 14\mbox{(syst)}$ MeV. Thus, a complementary measurement
by an indpendent experiment is required to resolve this controversy. Nevertheless, the statistical
significance of the observed signal is sufficient to claim the observation of this baryon
by both the D\O\ and CDF experiments.

\subsection{Excited $B$ mesons}

In addition to the ground state $B$ mesons with spin-parity $J^P = 0^-$ ($\Bpm$, $\Bd$, $\Bs$, $\Bcpm$) and $B^*$ mesons
$(J^P=1^-)$ with internal orbital momentum $L=0$,
the quark model predicts a rich spectrum of excited $B$ mesons with higher values of $L$.
In particular, for $L=1$, there should be the broad states $B^*_0$ ($J^P = 0^+$) and $B^*_1$ ($J^P = 1^+$), and
the narrow states $B_1$ ($J^P = 1^+$) and $B_2^*$ ($J^P = 2^+$).
\cite{Matsuki,Matsuki-1,DiPierro,Eichten,Isgur,Isgur-1,Isgur-2,Ebert1,Orsland,Falk}
The broad states decay through an $S$ wave and therefore have widths of a
few hundred MeV. Such states are difficult to distinguish in the invariant mass distribution from the combinatorial
background. The narrow states decay through a $D$ wave and therefore should have widths of around 10 MeV.
The masses, widths and the relative branching fractions of these states are predicted with
good precision by the theoretical models and can be compared with the experimental results.

Both collaborations preformed the measurements of the neutral narrow excited $B$ mesons.
The direct decay $B_1^0 \to B^+ \pi^-$ is forbidden by conservation of parity and angular momentum.
Therefore, this meson was searched for in the decay mode $B_1^0 \to B^{*+} \pi^-$ with $B^{*+} \to B^+ \gamma$.
On the contrary, both decays $B_2^{*0} \to B^{*+} \pi^-$ and $B_2^{*0} \to B^+ \pi^−$ are allowed, and
the corresponding branching fractions are almost equal.\cite{DiPierro,Eichten,Falk} Since the soft photon
from the decay $B^{*+} \to B^+ \gamma$ is not reconstructed, these decays produce three visible peaks in
the distribution of the mass difference $\Delta m = m(B^+ \pi^-) - m(B^+)$, with the signal from the decays
$B_1^0 \to B^{*+} \pi^-$ and $B_2^{*0} \to B^+ \pi^−$ shifted from the nominal position by
$\delta m = M(B^{*+}) - M(B^+) = 45.8$ MeV.

\begin{figure}[tpbh]
\begin{center}
\centerline{
\psfig{file=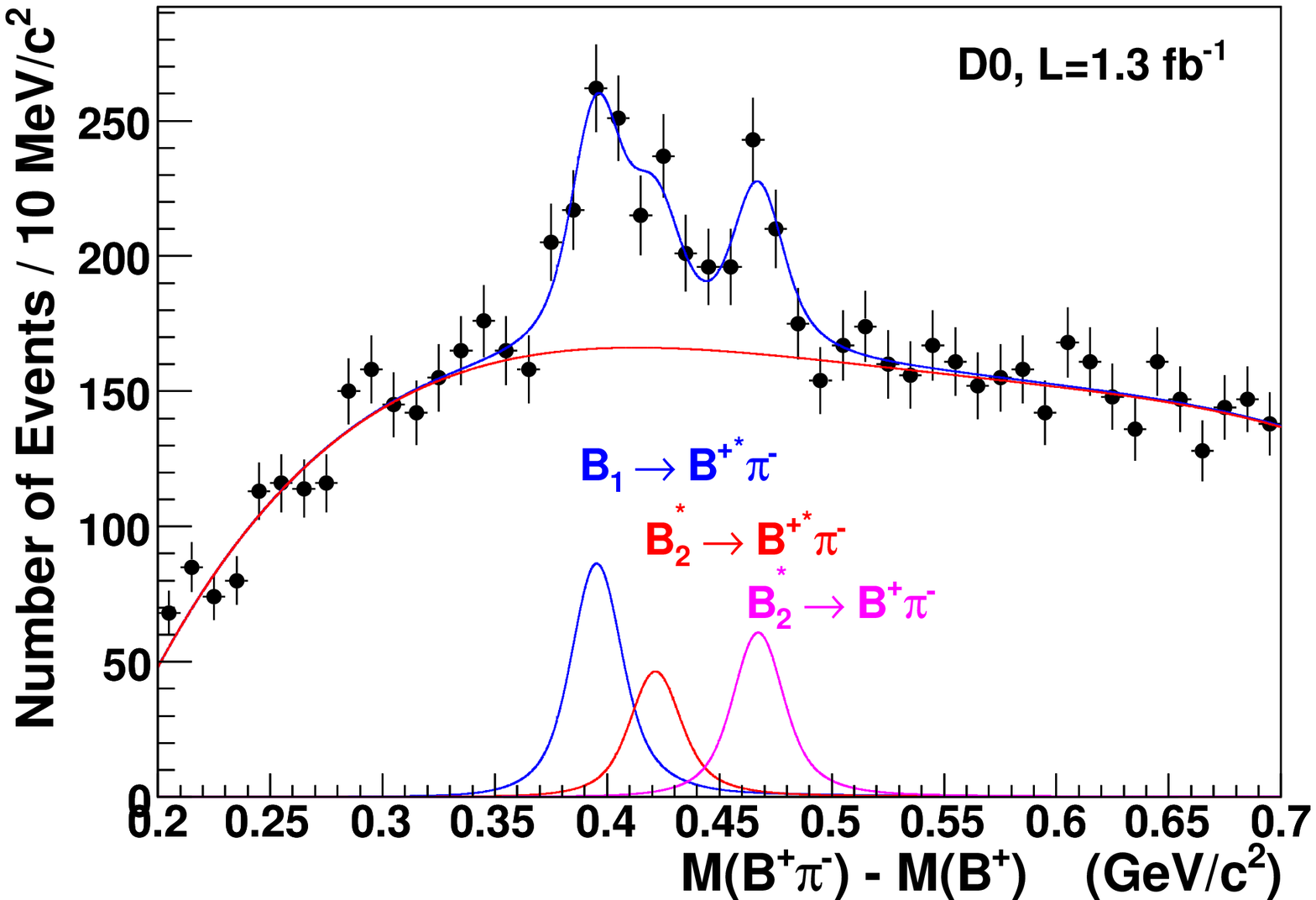,width=0.55\textwidth}
\psfig{file=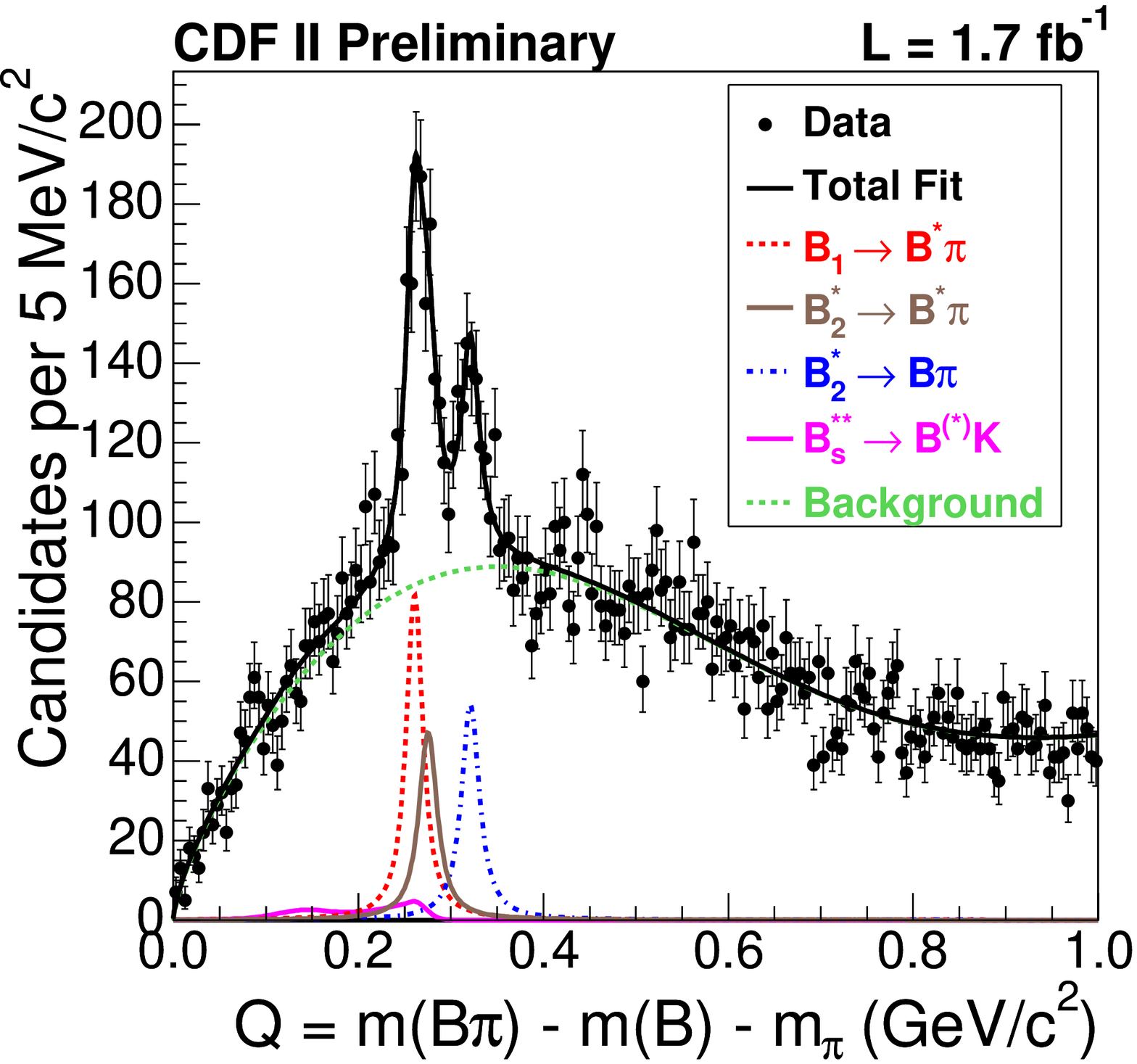,width=0.45\textwidth}
}
\vspace*{8pt}
\caption{
The left plot taken from Ref. \cite{bstst-d0} shows the invariant mass difference $m(B^+ \pi^-) - m(B^+)$
for the exclusive $B^+$ decays obtained
by D\O\ collaboration. The line shows the fit to the obtained distribution.
The contribution of background and the three signal peaks are shown separately.
The right plot taken from Ref. \cite{bstst-cdf} shows the distribution of the mass
difference $Q = m(B^+ \pi^-) - m(B^+) - m(\pi^-)$
for exclusive B+ decays obtained by the CDF collaboration.
Curves are shown separately for the background, the $B_s^{**} \to B^{(*)} K$ reflections,
and the three decays of excited $B$ mesons.
\label{bstst-mass}}
\end{center}
\end{figure}

The D\O\ collaboration performed the analysis using 1.3 fb$^{-1}$ of statistics.\cite{bstst-d0}
The $B^+$ meson was selected in the decay mode $B^+ \to J/\psi K^+$ with $J/\psi \to \mu^+ \mu^-$.
The obtained distribution
of the difference $m(B^+ \pi^-) - m(B^+)$ is shown in the left plot in Fig.~\ref{bstst-mass}.
The signal of the excited $B$ mesons is clearly seen, although the separation of the three decay modes
is quite difficult, and involves the theoretical assumption on the $B_1$ and $B_2^*$ decay widths.
The CDF collaboration used in their study\cite{bstst-cdf} three decay modes of $B^+$, namely
$B^+ \to J/\psi K^+$ with $J/\psi \to \mu^+ \mu^-$,
$B^+ \to \bar{D}^0 \pi^+$, and $B^+ \to \bar{D}^0 \pi^+ \pi^- \pi^+$ with $\bar{D}^0 \to K^+ \pi^-$.
The distribution of the $Q$ values defined as $Q = m(B^+ \pi^-) - m(B^+) - m(\pi^-)$ is shown
in the right plot in Fig.\ \ref{bstst-mass}. The signal of the excited $B$ mesons is also clearly
seen and in a qualitative agreement with the result obtained by the D\O\ collaboration.

The D\O\ collaboration reports the following measured masses of $B_1$ and $B_2^*$ mesons:
\begin{eqnarray}
m(B_1)   & = & 5720.6 \pm 2.4\mbox{(stat)} \pm 1.4\mbox{(syst)} ~\mbox{MeV}, \\
m(B_2^*) & = & 5746.8 \pm 2.4\mbox{(stat)} \pm 1.7\mbox{(syst)} ~\mbox{MeV}.
\end{eqnarray}
In addition, the following ratios were measured:
\begin{eqnarray}
\frac{{\rm Br}(B_1 \to B^{*} \pi)}{{\rm Br}(B_1, B_2^* \to B^{(*)} \pi)} & = & 0.477 \pm 0.069\mbox{(stat)} \pm 0.062\mbox{(syst)}; \\
\frac{{\rm Br}(B_2^* \to B^{*} \pi)}{{\rm Br}(B_2^* \to B^{(*)} \pi)} & = & 0.475 \pm 0.095\mbox{(stat)} \pm 0.069\mbox{(syst)}; \\
\frac{{\rm Br}(b \to (B_1, B_2^*) \to B^{(*)+} \pi^-)}{{\rm Br}(b \to B^+)} & = & 0.139 \pm 0.019\mbox{(stat)} \pm 0.032\mbox{(syst)}.
\end{eqnarray}

The CDF collaboration found the masses of $B_1$ and $B_2^*$ mesons to be equal to
\begin{eqnarray}
m(B_1)   & = & 5725.3^{+1.6}_{-2.2}\mbox{(stat)} ^{+1.4}_{-1.5}\mbox{(syst)} ~\mbox{MeV}, \\
m(B_2^*) & = & 5740.2^{+1.7}_{-1.8}\mbox{(stat)} ^{+0.9}_{-0.8}\mbox{(syst)} ~\mbox{MeV}.
\end{eqnarray}
In addition, they measured the natural width of the $B_2^*$ meson, which is found to be
\begin{equation}
\Gamma(B_2^{*0}) = 22.7^{+3.8}_{-3.2}\mbox{(stat)} ^{+3.2}_{-10.2}\mbox{(syst)} ~\mbox{MeV}.
\end{equation}
The masses measured by both experiments are in a reasonably good agreement between them and with the theoretical predictions.

\subsection{Excited $B_s$ mesons}
The $(b \bar{s})$ system should reveal the same pattern of the excited mesons as the $(b \bar{d})$ system.
The quark theory predicts the existence of two broad states $B_{s0}^*$ and $B_{s1}^*$, and
two narrow states $B_{s1}$ and $B_{s2}^*$. The narrow states can be observed experimentally.
The $B_{s1}$ meson can only decay to $B_{s1} \to B^* K$, while $B_{s2}^*$ meson can decay to both
$B_{s2}^* \to B^* K$ and $B_{s2}^* \to B K$ final states. However, the decay $B_{s2}^* \to B^* K$ is strongly
suppressed due to a small available phase space.

\begin{figure}[tpbh]
\begin{center}
\centerline{
\psfig{file=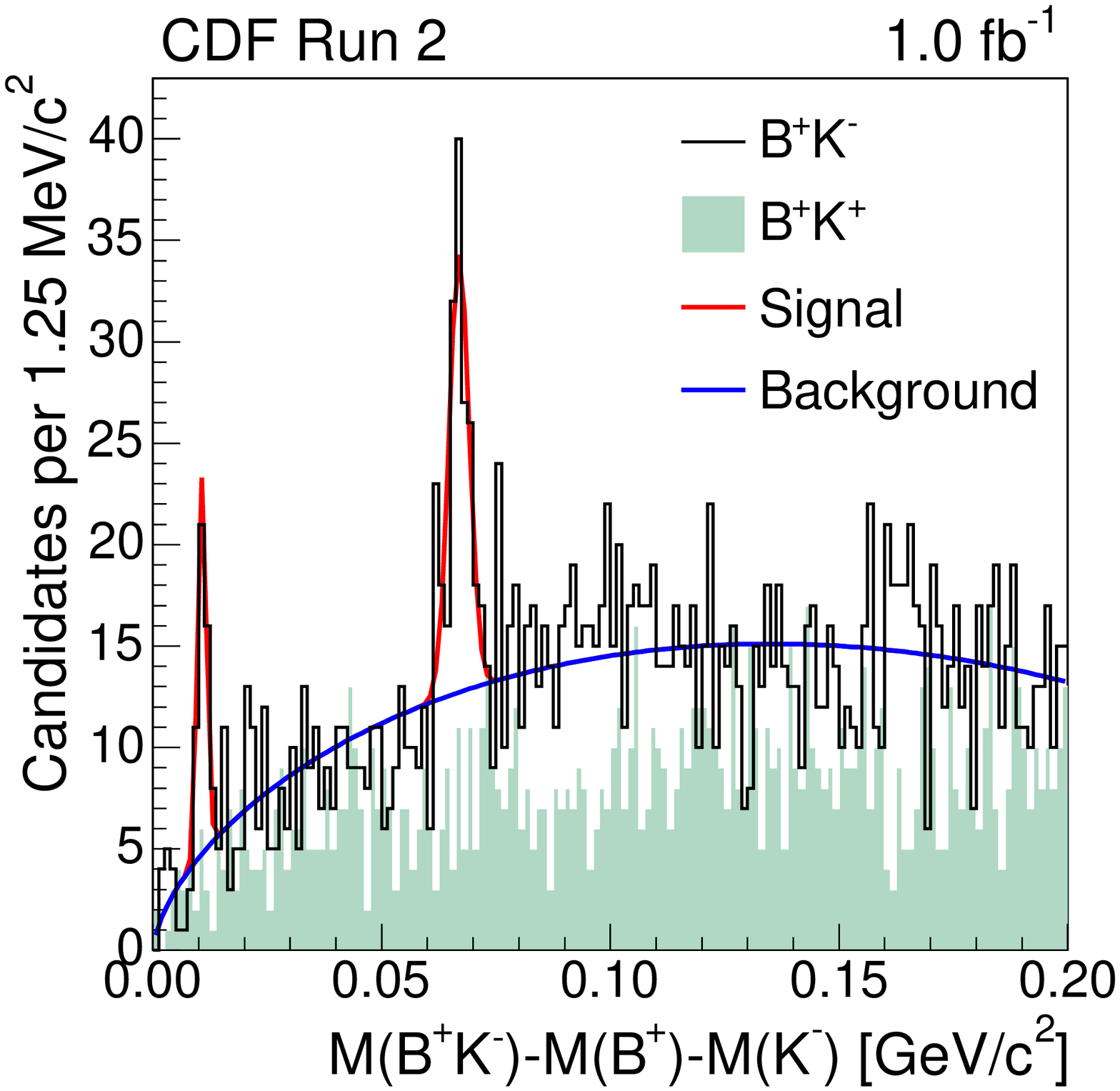,width=0.45\textwidth}
\psfig{file=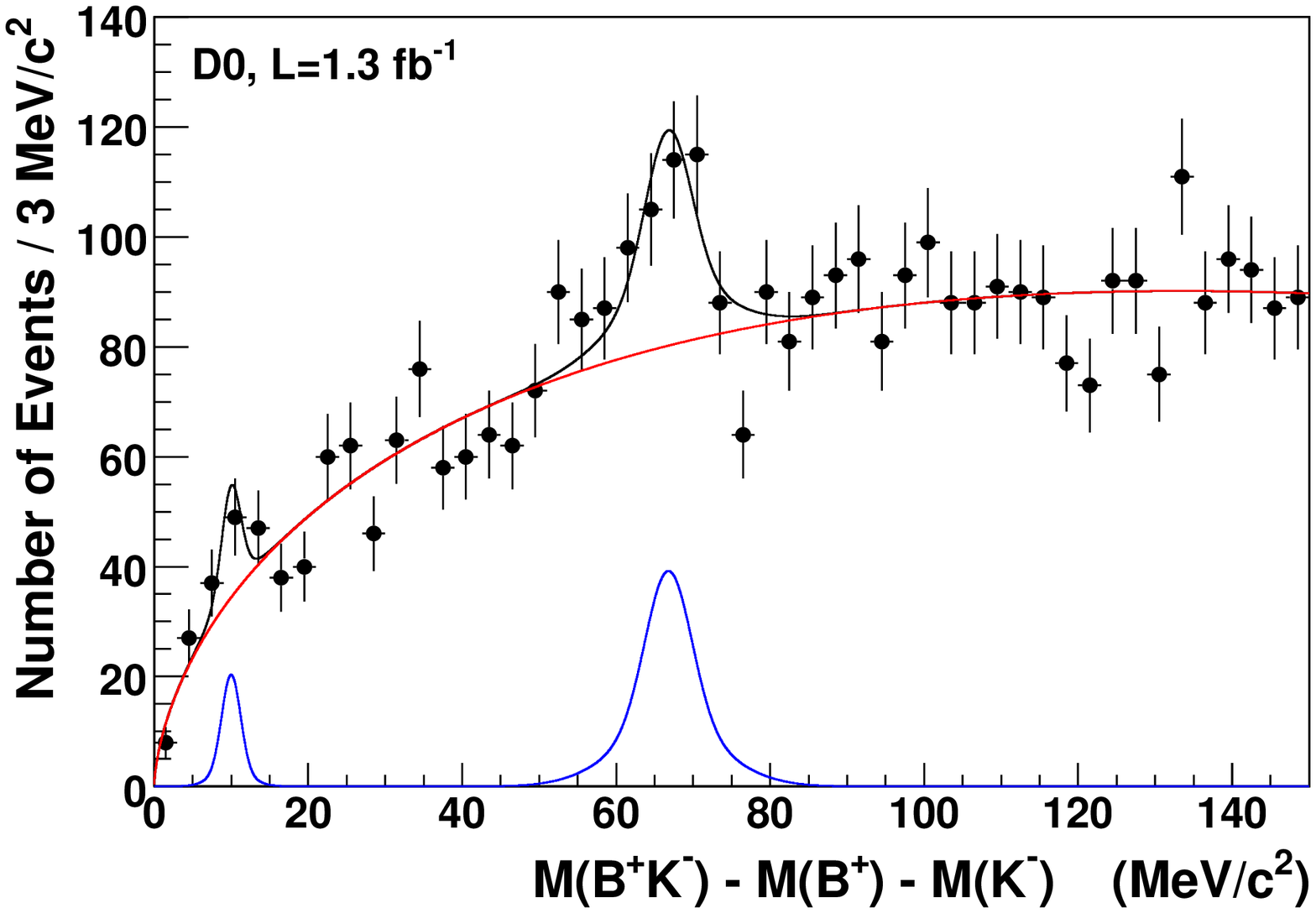,width=0.55\textwidth}
}
\vspace*{8pt}
\caption{
The left plot taken from Ref. \cite{bs-stst-cdf} shows the distribution of $Q = m(B^+ K^-) - m(B^+) - m(K-)$
obtained by the CDF collaboration for both $B^+$ channels combined. The dotted line shows
the result of a fit with the sum of a background function and
two Gaussians. The filled area shows the $Q$ distribution for the wrong-sign combination $B^+ K+$.
The right plot taken from Ref. \cite{bs-stst-d0} shows the invariant mass difference $m(B^+ K^-) - m(B^+) - m(K-)$
for exclusive $B^+$ decays obtained by the D\O\ collaboration. The line shows
the fit with a two-peak hypothesis. Shown separately are contributions from signal and background.
\label{bs-stst-mass}}
\end{center}
\end{figure}

The CDF collaboration reconstructed the $B^+$ meson in the decays modes
$B^+ \to J/\psi K^+$ with $J/\psi \to \mu^+ \mu^-$,
and $B^+ \to \bar{D}^0 \pi^+$ with $\bar{D}^0 \to K^+ \pi^-$.
They analysed\cite{bs-stst-cdf} the statistics collected with 1 fb$^{-1}$ of
$p \bar p$ collisions. The obtained distribution of the difference $m(B^+ K^-) - m(B^+) - m(K-)$ is shown in the
left plot in Fig. \ref{bs-stst-mass}. The signal corresponding to the decays $B_{s1} \to B^{*+} K^-$ and
$B_{s2}^* \to B^+ K^-$ are clearly seen. The statistical significance of each of the signals exceeds five
standard deviations. The result of the similar study\cite{bs-stst-d0} reported by the D\O\ collaboration is shown in the right plot
in Fig.\ \ref{bs-stst-mass}. They used the decay $B^+ \to J/\psi K^+$ with $J/\psi \to \mu^+ \mu^-$ in their analysis.
The D\O\ collaboration observes the signal of the $B_{s2}^* \to B^+ K^-$ decay with the statistical significance
exceeding 5 standard deviations, but they were not able to stated any conclusive statement on the
$B_{s1} \to B^{*+} K^-$ decay due to the insufficient significance of the possible signal.

The masses of $B_{s1}$ and $B_{s2}^*$ mesons measured by the CDF collaboration are found to be
\begin{eqnarray}
m(B_{s1})   & = & 5829.4 \pm 0.2\mbox{(stat)} \pm 0.1\mbox{(syst)} \pm 0.6\mbox{(PDG)} ~\mbox{MeV}, \\
m(B_{s2}^*) & = & 5839.6 \pm 0.4\mbox{(stat)} \pm 0.1\mbox{(syst)} \pm 0.5\mbox{(PDG)} ~\mbox{MeV}.
\end{eqnarray}
The last uncertainty in these results is due to the uncertainty of different values taken from the Particle
Data Group \cite{pdg}.

The mass of the $B_{s2}^*$ meson obtained by the D\O\ collaboration is
\begin{equation}
m(B_{s2}^*) = 5839.6 \pm 1.1\mbox{(stat)} \pm 0.7\mbox{(syst)} ~\mbox{MeV}.
\end{equation}
In addition, the D\O\ collaboration measured the production rate of $B_{s2}^*$ meson
\begin{equation}
\frac{{\rm Br}(b \to B_{s2}^* \to B^{+} K^-)}{{\rm Br}(b \to B^+)} =  (1.15 \pm 0.23\mbox{(stat)} \pm 0.13\mbox{(syst)}) \%.
\end{equation}
The results of two collaborations agree very well between them.

\subsection{Conclusions}
Concluding this section, it can be stated that the experiments at the Tevatron obtained the impressive results on the
spectroscopy of the $B$ hadrons. Many objects were observed for the first time and their measured parameters provide
an important input for improving and developing the theoretical models describing the bound states of quarks.
Still, many $B$ hadrons, like the baryons containing two or three $b$ quarks, need to be discovered, and
the experiments at the LHC collider have many possibilities to contribute in this exciting direction of $B$ physics.

\section{Lifetime of $B$ hadrons}
\label{lifetimes}

The lifetime is one of the most important properties of hadron. It is determined by
the interplay of the strong interaction which bounds quarks together,
and the weak interaction responsible for the transition of one quark type to another.
Therefore measuring the lifetime provides an important information on the underlying theory
of strong interactions (QCD) and helps to develop the numerical models used to predict the $B$-hadron lifetime.

A large mass of $b$ quark significantly simplifies the computation of the $B$-hadron lifetime, since
the QCD at the energy corresponding to $m(b) = 4.12 \pm 0.03$ GeV\cite{pdg-2012} becomes a perturbative theory,
so that its predictions become precise and unambiguous with small theoretical uncertainties. According
to this prediction, the lifetimes of stable $B$ hadrons should exhibit the following
striking hierarchy:
\begin{equation}
\tau(B^+) > \tau(\Bd) \approx \tau(\Bs) > \tau(\Lambda_b) \gg \tau(B_c)
\label{pattern}
\end{equation}
The lifetime of all $B$ hadrons, except the lifetime of $B_c$ meson, should be the same within 10\%,
while the lifetime of the $B_c$ meson should be considerably smaller. The experimental task
is to verify this prediction. Until the start of the LHC experiments, essentially all what was known
about the $B$-hadron lifetimes, except the lifetimes of $B^+$ and $\Bd$ mesons,
came from the Tevatron experiments.
And even for $B^+$ and $\Bd$ mesons, the CDF and D\O\ experiments provide an important contribution in the
world average values, because the accuracy achieved by them is comparable to the results of the
$b$-factory experiments.

The theoretical prediction is more precise for the ratio of lifetimes of two $B$ hadrons. For example,
the ratio of $\Bs$ and $\Bd$ lifetimes is predicted with the theoretical uncertainty of the order of 1\%.
Form the experimental point of view, the measurement of the ratio of lifetimes is also often more precise
than the measurement of the individual lifetimes, because many systematic uncertainties cancel
in the corresponding ratio. Therefore, the comparison of the theoretical and experimental results
on the lifetime ratios is the most meaningful and educative, and in this section we will pay a special
attention to the experimental results on the lifetime ratios.

\subsection{$B^+$ and $\Bd$ lifetimes}

\begin{figure}[tpbh]
\begin{center}
\centerline{
\psfig{file=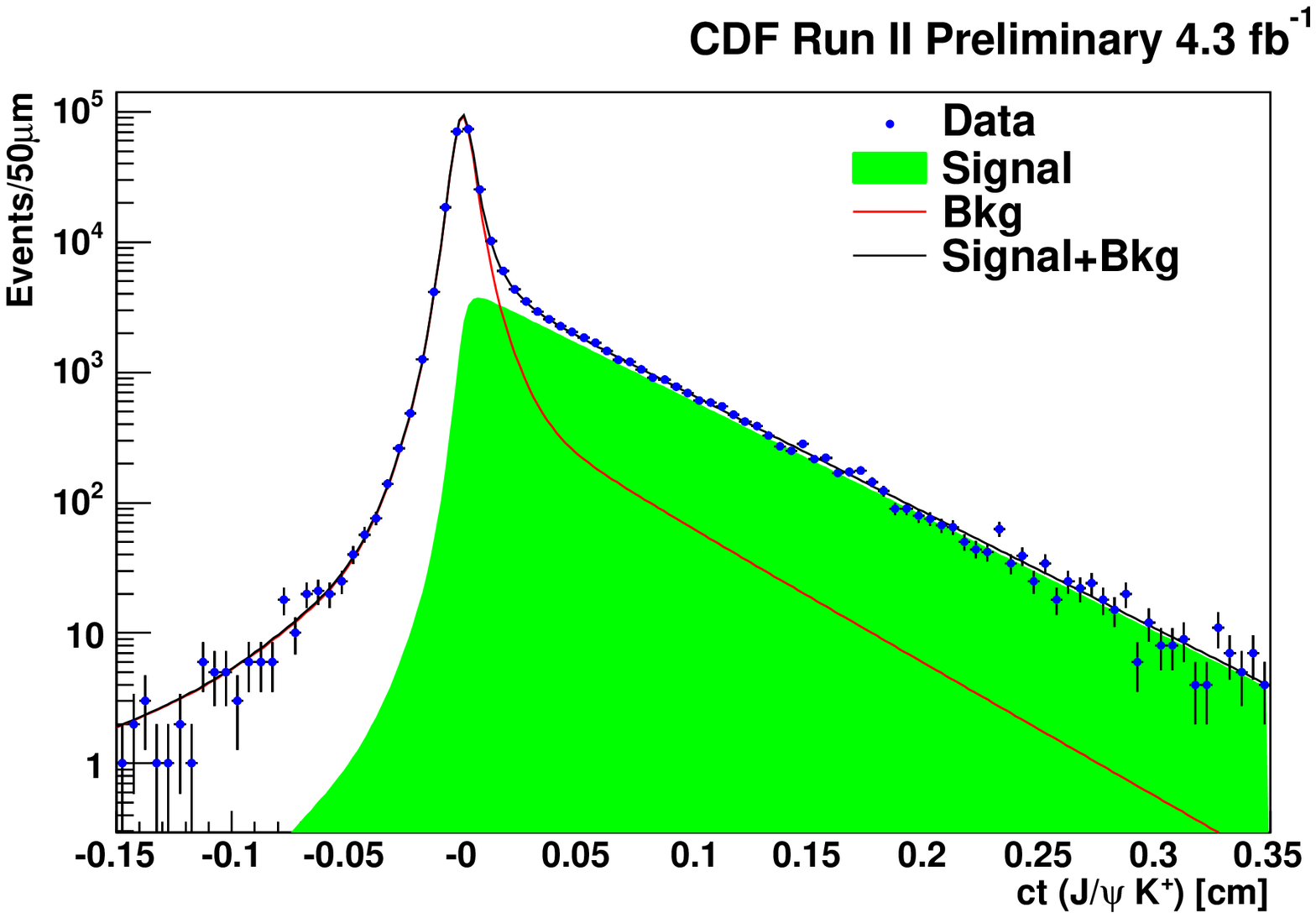,width=0.50\textwidth}
\psfig{file=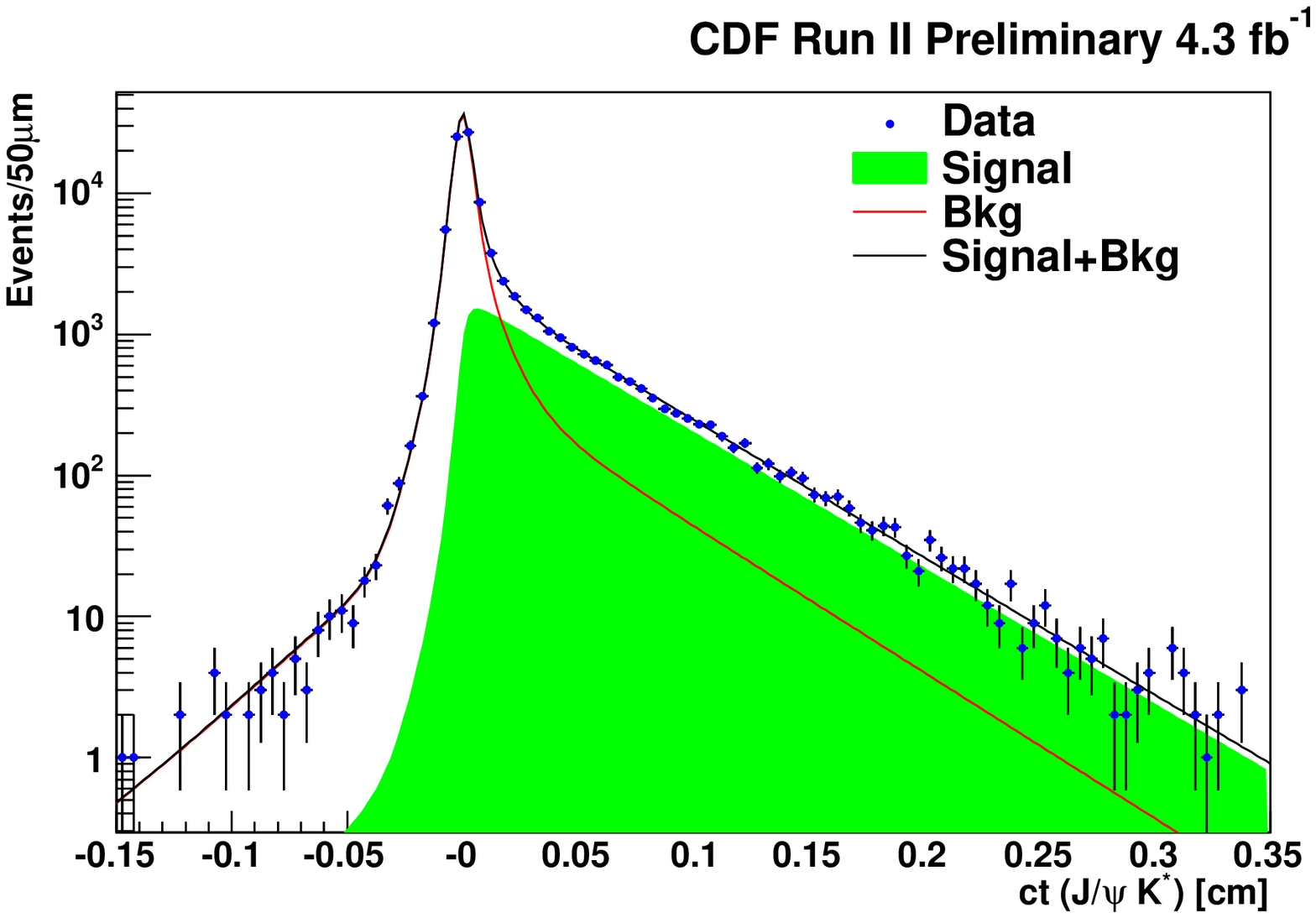,width=0.50\textwidth}
}
\vspace*{8pt}
\caption{
Decay time distributions for $B^+ \to J/\psi K^+$ (left plot) and
$\Bd \to J/\psi K^{*0}$ (right plot) candidates taken from Ref. \cite{b-lifetime-jpsi-cdf}.
The decay time projection of the likelihood function is shown overlaid.
\label{b-lifetime-jpsi}}
\end{center}
\end{figure}

The CDF collaboration performed several measurements of the $B^+$ and $\Bd$ lifetimes. The most precise result
comes from the study of $B$ hadron decays involving
$J/\psi \to \mu^+ \mu^-$ final state.\cite{b-lifetime-jpsi-cdf} This final
state can be selected using the di-muon trigger without applying the lifetime biasing cuts. Using
the statistics corresponding to an integrated luminosity of 4.3 fb$^{-1}$, the CDF collaboration collected
$43000 \pm 230$ decays $B^+ \to J/\psi K^+$, $16860 \pm 140$ decays $\Bd \to J/\psi K^{*0}$ and
$12070 \pm 120$ decays $\Bd \to J/\psi K_s^0$. For each collected sample
the lifetime of $B^+$ and $\Bd$ mesons is obtained using an unbinned maximum likelihood fit.
The corresponding likelihood function {\cal L} is multivariate,
and is based on the probability of observing a given candidate with reconstructed mass, decay time,
decay time uncertainty, and mass uncertainty. The comparison of the decay time distribution
and the projections of the likelihood function after the fit is shown in Fig.\ \ref{b-lifetime-jpsi}.
It can be seen that the quality of the description of data is good.

Using the collected statistics, the CDF collaboration obtained the following results:
\begin{eqnarray}
\tau(B^+) & = & 1.639 \pm 0.009\mbox{(stat)} \pm 0.009\mbox{(syst)} ~\mbox{ps}, \\
\tau(\Bd) & = & 1.507 \pm 0.010\mbox{(stat)} \pm 0.008\mbox{(syst)} ~\mbox{ps}.
\end{eqnarray}
These results are consistent and have the similar precision as the corresponding measurement by Belle
collaboration.\cite{b-tau-belle}
The ratio of lifetimes of $B^+$ and $\Bd$ mesons provides an interesting possibility of comparison with
the theoretical prediction, which is very precise due to the cancelation of many theoretical
uncertainties. On the other hand, the systematic uncertainty of the experimental result is also reduced
in the ratio of the lifetimes. The ratio of lifetimes $\tau(B^+) / \tau(\Bd)$ found
by the CDF collaboration is
\begin{equation}
\tau(B^+) / \tau(\Bd) = 1.088 \pm 0.009\mbox{(stat)} \pm 0.004\mbox{(syst)}.
\end{equation}
The theoretical prediction\cite{Beneke,Beneke-1,Beneke-2,Beneke-3,Bigi,Gabbiani,Gabbiani-1} 
of this ratio is in the range $1.04 - 1.08$,
which is consistent with this experimental ratio.

\begin{figure}[tpbh]
\begin{center}
\centerline{
\psfig{file=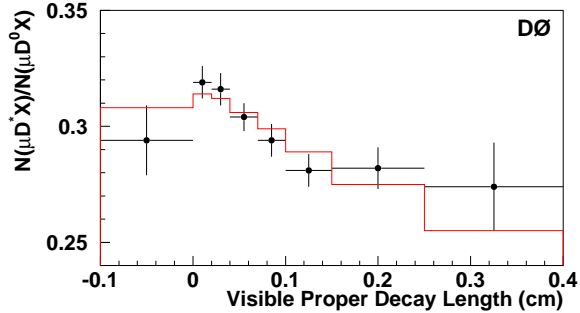,width=0.70\textwidth}
}
\vspace*{8pt}
\caption{
Points with the error bars show the ratio of the number
of events in the $\mu^+ D^{*-}$  and $\mu^+ \bar{D}^0$  samples as a function
of the visible proper decay length. The result of the fit of this distribution with
$\tau(B^+) / \tau(\Bd) = 0.080$ is shown as a histogram. The plot is taken from Ref. \cite{b-tau-d0}.
\label{b-tau-ratio-d0}}
\end{center}
\end{figure}

The D\O\ collaboration performed an original direct measurement\cite{b-tau-d0} of the ratio of the
lifetimes of $B^+$ and $\Bd$ mesons
using the semileptonic decays $B^+ \to \mu^+ \nu \bar{D}^0 X$ and
$\Bd \to \mu^+ \nu D^{*-}$ with $D^{*-} \to \bar{D}^0 \pi$
and $\bar{D}^0 \to K^+ \pi^-$.
Using just 440 pb$^{-1}$ of $p \bar p$ collisions,
the D\O\ collaboration collected more than 120000 events containing $\mu^+ \bar{D}^0$ final state.
Since the neutrino in the decays of $B$ mesons is not reconstructed,
the proper decay length can not be determined. Therefore, the D\O\ collaboration defined the
visible proper decay length ($VPDL$) using the measured decay length in the laboratory frame
and the total reconstructed momentum of the $\mu^+ \bar{D}^0$ or $\mu^+ D^{*-}$ system.
They measured the ratio $R$ of the number of $\mu^+ D^{*-}$  and $\mu^+ \bar{D}^0$ events
as a function of the $VPDL$.
If the lifetime of the $B^+$ and $\Bd$ mesons is different, the ratio $R$ should change with the change
of the $VPDL$. Namely this variation is observed experimentally, as it can be seen
in Fig.\ \ref{b-tau-ratio-d0}, and from the shape of this variation the ratio $\tau(B^+) / \tau(\Bd)$
can be extracted. Since the final states $\mu^+ \bar{D}^0$ or $\mu^+ D^{*-}$ are very similar
and can be selected using almost the same criteria, many
experimental uncertainties cancel in this $\tau(B^+) / \tau(\Bd)$ measurement.
Using this method, the D\O\ collaboration obtained the value
\begin{equation}
\tau(B^+) / \tau(\Bd) = 1.080 \pm 0.016\mbox{(stat)} \pm 0.014\mbox{(syst)}.
\end{equation}
This result agrees with the theoretical expectations and with the other measurements of this quantity.

\subsection{$\Bs$ lifetime}

$\Bs$ meson system, like any other neutral meson, contains the short-lived and and long-lived states.
They differ by mass and lifetime and are denoted as the light ($B_L$) and heavy $(B_H)$ states, respectively,
with the value of $\Delta \Gamma = \Gamma_L - \Gamma_H$ to be positive in the SM.
Since the value of $\Delta \Gamma_s$ is relatively large, the measured lifetime of the $B_s$ system
depends on the $\Bs$ decay mode in which it is measured. The distinctive cases are the flavor
specific final state, the $\Bs \to J/\psi \phi$ decay, which is a mixture of $CP$-even
and $CP$-odd final state, and $CP$-specific final state, like the $\Bs \to J/\psi f_0(980)$ decay.
The Tevatron collaborations contributed in the measurement of the $\Bs$ lifetime in all these final states,
and the precision of their measurements was the world best until the start of the LHC experiments.

\begin{figure}[tpbh]
\begin{center}
\centerline{
\psfig{file=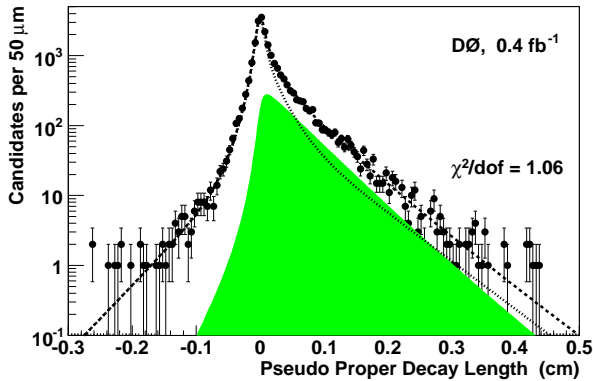,width=0.70\textwidth}
}
\vspace*{8pt}
\caption{
Pseudo-proper decay length distribution for $\mu^+ D_s^-$
candidates with the result of the fit superimposed as the
dashed curve. The dotted curve represents the combinatorial
background and the filled area represents the $\Bs$ signal. The plot is taken from Ref. \cite{bs-fs-d0}.
\label{bs-tau-fs-d0}}
\end{center}
\end{figure}

The flavour specific lifetime of the $\Bs$ system is measured by the D\O\ collaboration\cite{bs-fs-d0}
in the semileptonic decay $\Bs \to \mu^+ D_s^- X$. Using just 400 pb$^{-1}$ of available statistics,
they reconstructed about 5000 decays $\Bs \to \mu^+ D_s^- X$ and measured the $\Bs$ lifetime to be
\begin{equation}
\tau(\Bs \to \mu^+ D_s^- X)  =  1.398 \pm 0.044\mbox{(stat)} ^{+0.028}_{-0.025}\mbox{(syst)} ~\mbox{ps}.
\end{equation}
Figure\ \ref{bs-tau-fs-d0} shows the distribution of the pseudo proper decay length, which
is another name of VPDL introduced before, of the reconstructed $\Bs \to \mu^+ D_s^- X$ decays together
with the result of the log likelihood fit superimposed. In general, there is an excellent description
of the observed data with the $\chi^2$ per degree of freedom ($dof$) $\chi^2/dof = 1.06$.
This result has a potential of considerable improvement, since less than 5\% of the final statistics
is used in the analysis, but the D\O\ collaboration never
updated it. Nevertheless, even with such a small statistics this result remains one of the
most precise flavour specific measurements of the $\Bs$ lifetime.

\begin{figure}[tpbh]
\begin{center}
\centerline{
\psfig{file=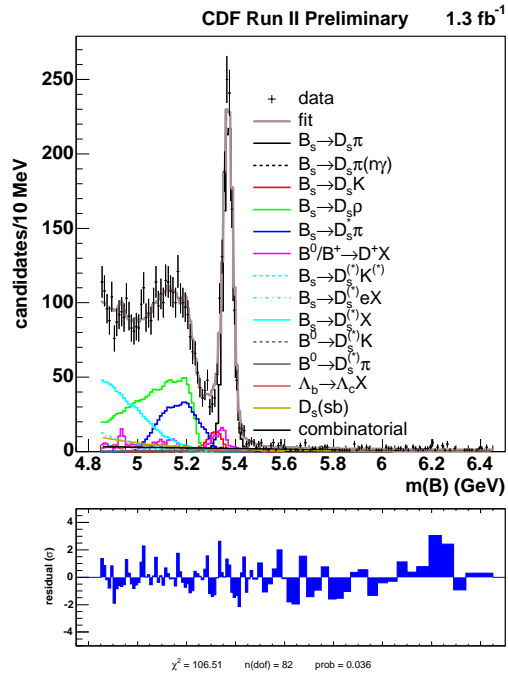,width=0.60\textwidth}
}
\vspace*{8pt}
\caption{
Invariant mass distribution for candidates reconstructed as
$\Bs \to D_s^- \pi^+$ with fit projection overlaid. The plot is taken from Ref. \cite{bs-fs-cdf}.
\label{bs-tau-fs-cdf-mass}}
\end{center}
\end{figure}

The CDF collaboration measured\cite{bs-fs-cdf} the flavor-specific lifetime of $\Bs$ meson using the decay
$\Bs \to D_s^- \pi^+ (X)$ with $D_s^- \to \phi \pi^-$ and X representing possible additional particles
which are not reconstructed. The statistics used in this analysis corresponds to 1.3 fb$^{-1}$ integrated
luminosity of $p \bar p$ collisions. Fig.\ \ref{bs-tau-fs-cdf-mass} shows the invariant mass dstribution
of selected $D_s^- \pi^+$ candidates. The CDF collaboration includes in the measurement
the exclusive decays of $\Bs$, seen in Fig.\ \ref{bs-tau-fs-cdf-mass} as the narrow peak, and
partially reconstructed $\Bs$ decays, seen as a bump on the left side from the peak. This addition
allowed to more than double the number of $\Bs$ decays used in the analysis, although adding some
difficulties in treatment of the partially reconstructed $\Bs$ decays. It can be seen from
Fig.\ \ref{bs-tau-fs-cdf-mass} that there is a very good understanding of the observed invariant
mass distribution and all essential decay modes of $B$ hadrons are included in the analysis.

Using the selected sample of $\Bs$ decays, the CDF collaboration obtained the following flavour specific
$\Bs$ lifetime:
\begin{equation}
\tau(\Bs \to D_s^- \pi^+ (X))  =  1.518 \pm 0.041\mbox{(stat)} \pm 0.027 \mbox{(syst)} ~\mbox{ps}.
\end{equation}

The $\Bs$ lifetime in the decay mode $\Bs \to J/\psi \phi$ is obtained from the full angular analysis
including the measurement of the lifetime difference $\Delta \Gamma_s$ and the possible $CP$-violating
phase $\phi_s$ and is described in Section \ref{cpv} of this review.

The CDF collaboration
also performed the measurement\cite{bs-f0-cdf} of the $\Bs$ lifetime in the pure $CP$-odd final state
$\Bs \to J/\psi f_0(980)$. Neglecting the $CP$ violation, which is predicted to be
very small in the SM, the $\Bs$ lifetime in this mode should correspond
to the lifetime of the $B_s^H$ state. The CDF collaboration reconstructed $502 \pm 37$ such decays
and used this statistics to measure the $\Bs$ lifetime:
\begin{equation}
\tau(\Bs \to J/\psi f_0(980))  =  1.70 ^{+0.12}_{-0.11}\mbox{(stat)} \pm 0.03 \mbox{(syst)} ~\mbox{ps}.
\end{equation}
At the time of the publication, it was the first measurement of the $\Bs$ lifetime in a decay
to $CP$ eigenstate.

\subsection{$\Bcm$ lifetime}

The lifetime of $\Bcm$ meson should be much less than the lifetime of all other
$B$ mesons because both $b$ and $\bar{c}$ quarks in $\Bcm$ meson decay weakly and, in addition,
these spectator quarks can annihilate. Within different theoretical approaches,
the $\Bcm$ lifetime is predicted to be\cite{Beneke1,Kiselev,Kiselev-1,Chang} in the range $0.36 - 0.53$ ps.
Currently, all what we know about the $\Bcm$ lifetime comes from the Tevatron experiments.

\begin{figure}[tpbh]
\begin{center}
\centerline{
\psfig{file=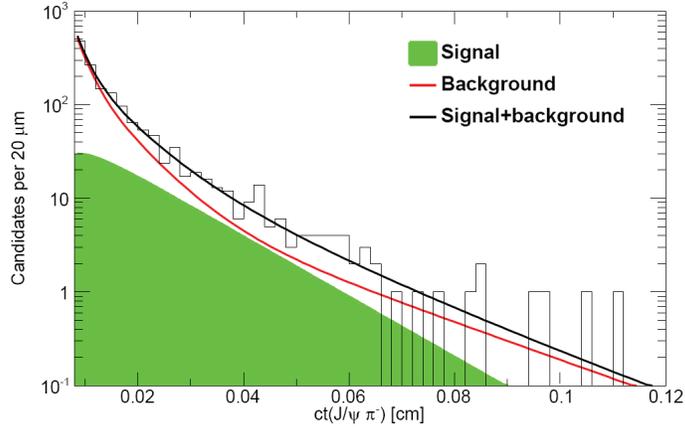,width=0.80\textwidth}
}
\vspace*{8pt}
\caption{
Decay-length distribution of $J/\psi \pi^-$ candidates taken from Ref. \cite{bc-tau-excl-cdf}. The
fit projection, along individual contributions from signal and
background, is overlaid.
\label{bc-tau-cdf}}
\end{center}
\end{figure}

The CDF collaboration measured\cite{bc-tau-sl-cdf} the $\Bcm$ lifetime using its semileptonic
decay $\Bcm \to J/\psi l^- \bar{\nu}$, where $l^-$ can be either an electron or a muon and $J/\psi \to \mu^+ \mu^-$.
They obtained
\begin{equation}
\tau(\Bcm)  =  0.475 ^{+0.053}_{-0.049}\mbox{(stat)} \pm 0.018 \mbox{(syst)} ~\mbox{ps}.
\end{equation}
In addition, the CDF collaboration measured\cite{bc-tau-excl-cdf} the $\Bcm$ lifetime in the exclusive decay mode
$\Bcm \to J/\psi \pi^-$ with $J/\psi \to \mu^+ \mu^-$. They reconstructed 272 exclusive $\Bcm$ decays
using the integrated luminosity 6.7 fb$^{-1}$. Figure \ref{bc-tau-cdf} shows the
proper decay length distribution of selected $\Bcm \to J/\psi \pi^-$ candidates with the result of the fit
superimposed. Using this sample of events, the CDF collaboration measured
\begin{equation}
\tau(\Bcm)  =  0.452 \pm 0.048\mbox{(stat)} \pm 0.027 \mbox{(syst)} ~\mbox{ps}.
\end{equation}

\begin{figure}[tpbh]
\begin{center}
\centerline{
\psfig{file=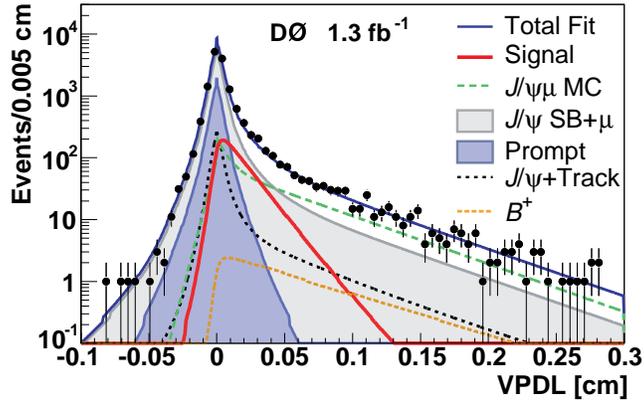,width=0.70\textwidth}
}
\vspace*{8pt}
\caption{
$VPDL$ distribution of the $J/\psi \mu$ sample with the projected
components of the fit overlaid. The plot is taken from Ref. \cite{bc-tau-sl-d0}.
\label{bc-tau-d0}}
\end{center}
\end{figure}

The D\O\ collaboration measured\cite{bc-tau-sl-d0} the $\Bcm$ lifetime using the semileptonic decay
$\Bcm \to J/\psi \mu^- \bar{\nu}$ with $J/\psi \to \mu^+ \mu^-$. They used the
statistics corresponding to the integrated luminosity 1.3 fb$^{-1}$ and reconstructed
$881 \pm 80$ $\Bcm$ candidates. The $VPDL$ distribution of selected candidates is shown
in Fig.\ \ref{bc-tau-d0}. The resulting $\Bcm$ lifetime is found to be
\begin{equation}
\tau(\Bcm)  =  0.448 ^{+0.038}_{-0.036}\mbox{(stat)} \pm 0.032 \mbox{(syst)} ~\mbox{ps}.
\end{equation}

All results on $\Bcm$ lifetime obtained by both collaboration are consistent with each other and
with the theoretical predictions\cite{Beneke1,Kiselev,Chang}.

\subsection{Lifetime of $B$ baryons}

Study of the $B$-baryon lifetime provides a valuable information for development of the numerical
theoretical models describing the quark systems. The lifetime of $\Lb$ baryon, containing ($bud$) quarks
is predicted\cite{Gabbiani,Tarantino} to be less than the lifetime of
$\Bd$ meson with $\tau(\Lb) / \tau(\Bd) = 0.92 \pm 0.03$. The early measurements\cite{pdg-2012} of $\Lb$ lifetime
from LEP were much less than the theoretical predictions, which explained an increased
interest to the $\Lb$ lifetime.

\begin{figure}[tpbh]
\begin{center}
\centerline{
\psfig{file=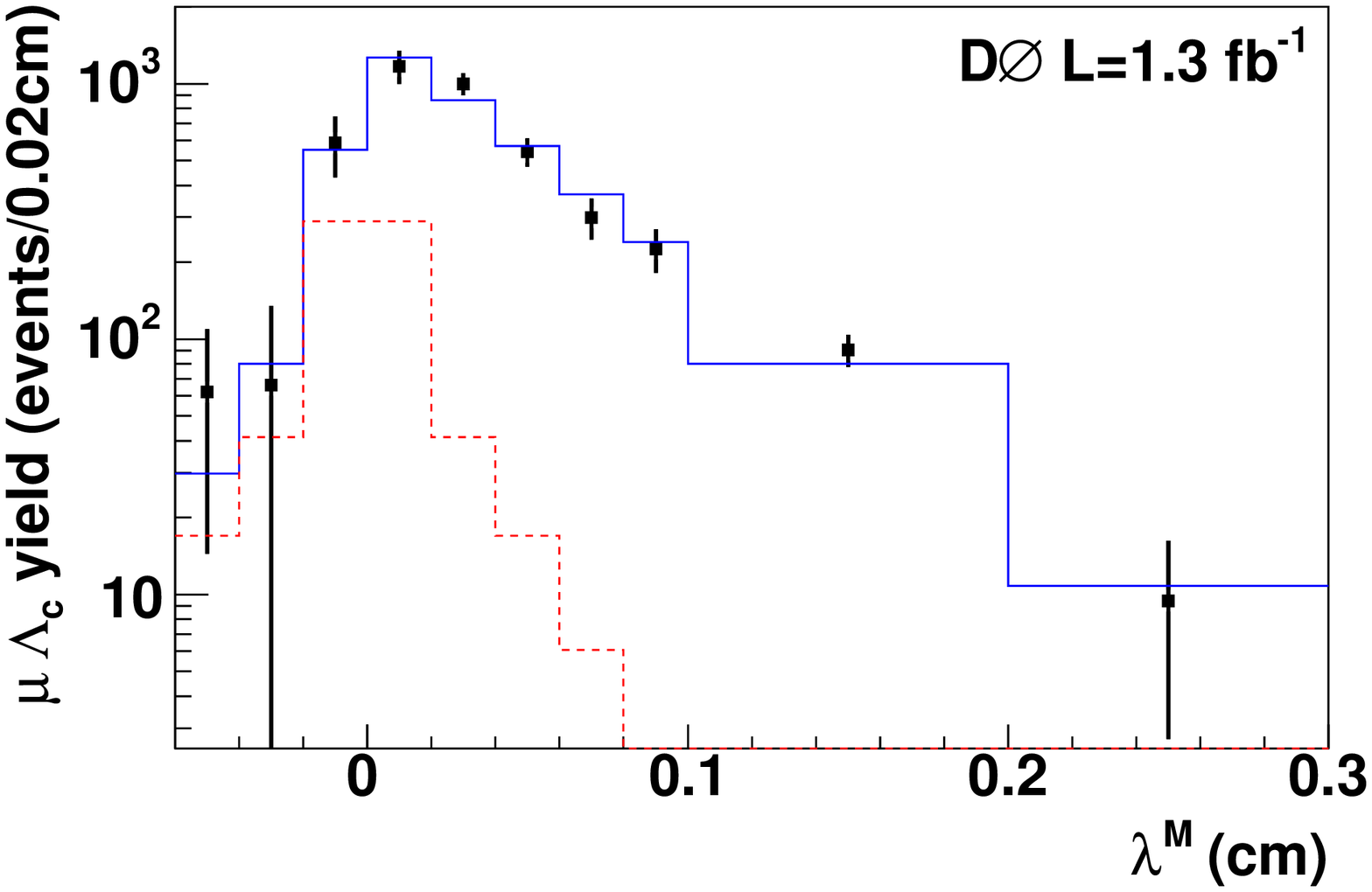,width=0.50\textwidth}
\psfig{file=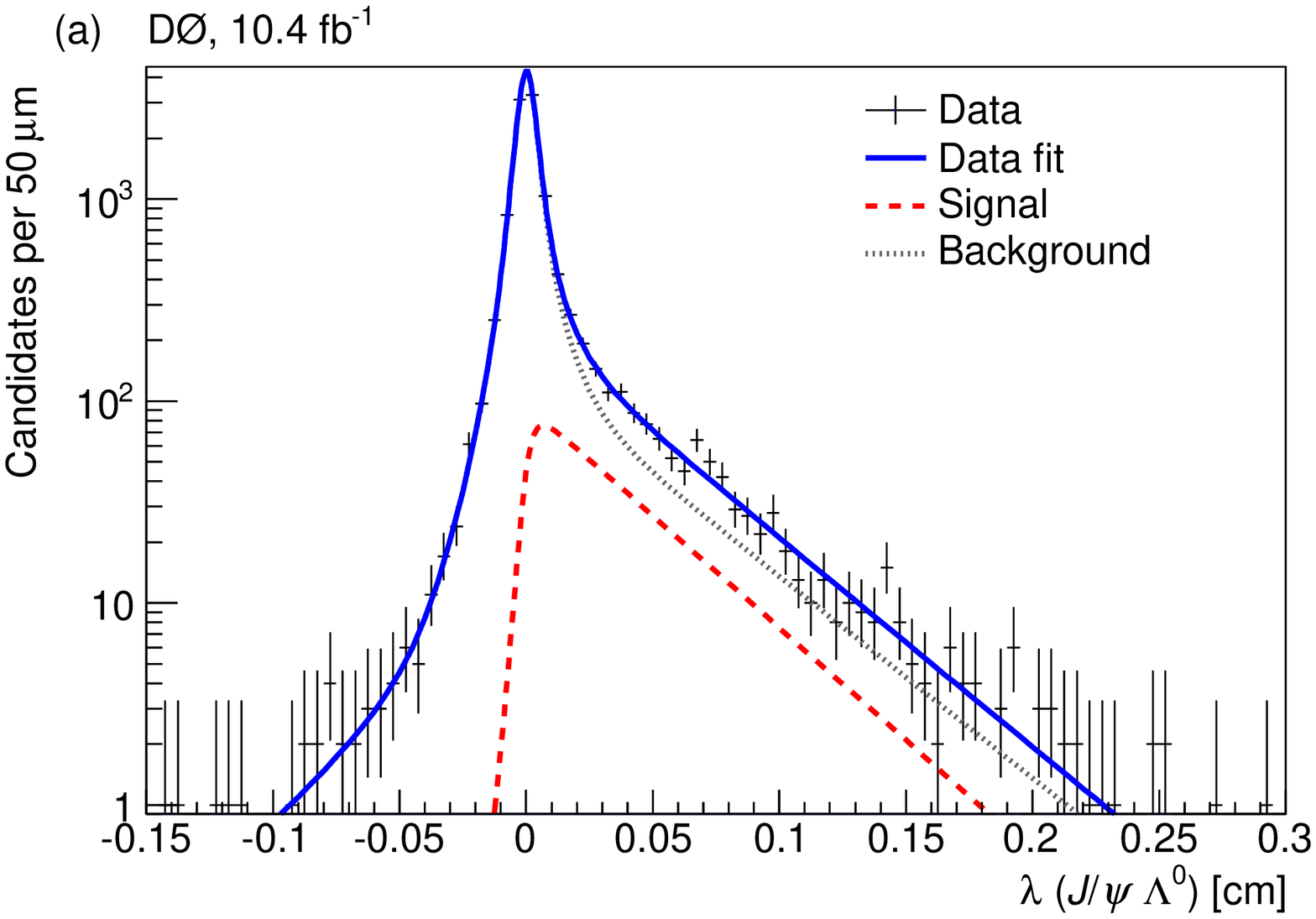,width=0.50\textwidth}
}
\vspace*{8pt}
\caption{
Left plot taken from Ref. \cite{lb-sl-d0}: Measured $\mu \Lambda_c$ yields in the $VPDL$ bins
denoted as $\lambda^M$ (points) and
the result of the lifetime fit (solid histogram). The dashed
histogram shows the contribution of peaking background. Right plot taken from Ref.\ \cite{lb-excl-d0}:
Proper decay length distributions for $\Lb \to J/\psi \Lambda$
candidates, with fit results
superimposed.
\label{lb-d0}}
\end{center}
\end{figure}

The D\O\ collaboration measured the $\Lb$ lifetime both in the semileptonic and hadronic decay modes.
For the first measurement\cite{lb-sl-d0} the semileptonic decay $\Lb \to \mu^- \bar{\nu} \Lambda_c^+$ is used with
$\Lambda_c^+ \to \Ks p$. In total $4437 \pm 329$ such decays are reconstructed
using the statistics corresponding to the integrated luminosity 1.3 fb$^{-1}$. The left plot in
Fig.\ \ref{lb-d0} shows the observed $VPDL$ distribution (denoted in Ref. \cite{lb-sl-d0} as $\lambda^M$)
together with the result of the fit. The $\Lb$ lifetime is found to be
\begin{equation}
\tau(\Lb)  =  1.290 ^{+0.119}_{-0.110}\mbox{(stat)} ^{+0.087}_{-0.091} \mbox{(syst)} ~\mbox{ps}.
\end{equation}

In the second measurement\cite{lb-excl-d0} of the $\Lb$ lifetime by the D\O\ collaboration, the decay
$\Lb \to J/\psi \Lambda$ with $J/\psi \to \mu^+ \mu^-$ and $\Lambda \to K^- p$ was used.
In total $755 \pm 49$ such decays were reconstructed using the full collected statistics corresponding
to the integrated luminosity 10.4 fb$^{-1}$. The proper decay length distribution of selected candidates
is shown in Fig.\ \ref{lb-d0} (right plot). From this statistics the measured $\Lb$ lifetime is found to be
\begin{equation}
\tau(\Lb)  =  1.303 \pm 0.075\mbox{(stat)} \pm 0.035 \mbox{(syst)} ~\mbox{ps}.
\end{equation}

The most precise measurements of the $\Lb$ lifetime by CDF collaboration were performed in hadronic
decay modes. The analysis in Ref. \cite{lb-lcp-cdf} exploits the decay $\Lb \to \Lambda_c^+ \pi^-$
with $\Lambda_c^+ \to p K^+ \pi^-$. Like in all other measurements with hadronic decays of $B$ hadrons,
the special trigger on the displaced vertex developed by the CDF collaboration is used.
Since such a trigger biases the lifetime distribution, a special attention in this analysis was
given to a correct reconstruction of the selection efficiency of $\Lb$ candidates. It is obtained
from the simulation of the trigger and detector, and is validated using $J/\psi \to \mu^+ \mu^-$ decays.

\begin{figure}[tpbh]
\begin{center}
\centerline{
\psfig{file=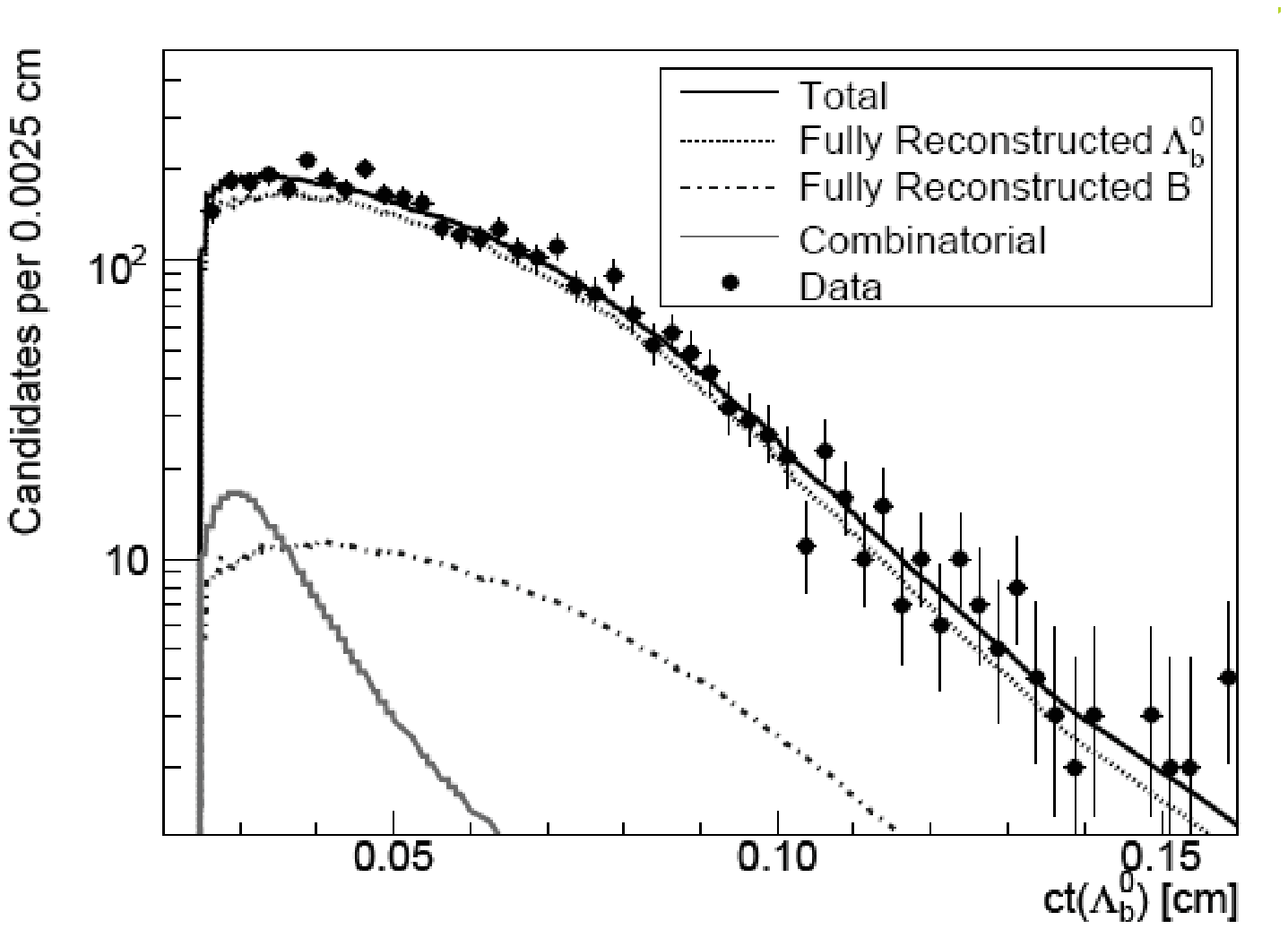,width=0.50\textwidth}
\psfig{file=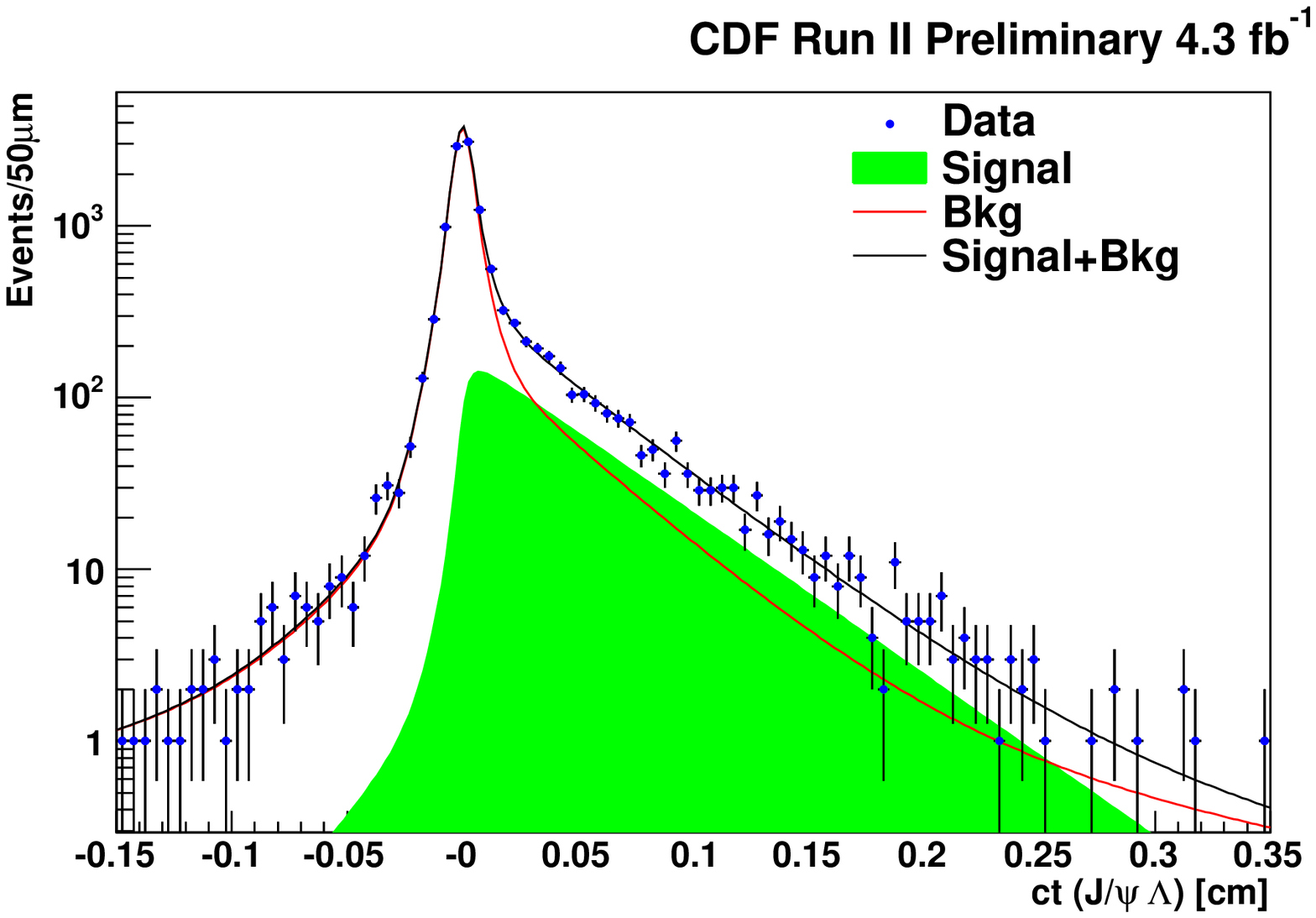,width=0.60\textwidth}
}
\vspace*{8pt}
\caption{
Left plot taken from Ref. \cite{lb-lcp-cdf}:
The distribution of the proper decay length ($c\tau$) of $\Lb \to \Lambda_c^+ \pi^-$ candidates
(points) with the fit projection overlaid (solid black line).
Right plot taken from Ref. \cite{b-lifetime-jpsi-cdf}:
The proper decay length ($c \tau$) distribution for $\Lb \to J/\psi \Lambda$ candidates.
\label{lb-cdf}}
\end{center}
\end{figure}

In total $2905 \pm 58$ decays $\Lb \to \Lambda_c^+ \pi^-$ were reconstructed using the full
collected statistics corresponding to the integrated luminosity 1.1 fb$^{-1}$. The reconstructed
proper decay length of $\Lb$ candidates together with the result of the fit
is shown in Fig.\ \ref{lb-cdf} (left plot). It can be seen that a very good description of
the reconstructed data, including a complicated trigger efficiency is achieved.
The measured $\Lb$ lifetime is
\begin{equation}
\tau(\Lb)  =  1.401 \pm 0.046\mbox{(stat)} \pm 0.035 \mbox{(syst)} ~\mbox{ps}.
\end{equation}

Another measurement\cite{b-lifetime-jpsi-cdf} of $\Lb$ lifetime by the CDF collaboration
is done in the decay mode
$\Lb \to J/\psi \Lambda$ with $J/\psi \to \mu^+ \mu^-$ and $\Lambda \to K^+ \pi^-$.
This analysis uses the statistics corresponding to the integrated luminosity of 4.3 fb$^{-1}$.
The analysis technique in this decay mode is straightforward and similar to the corresponding
measurement\cite{lb-excl-d0} by the D\O\ collaboration. In total $1710 \pm 50$ $\Lb$ decays were
reconstructed. The proper decay length distribution is shown in Fig.\ \ref{lb-cdf} (right plot).
The measured $\Lb$ lifetime is found to be
\begin{equation}
\tau(\Lb)  =  1.537 \pm 0.045\mbox{(stat)} \pm 0.014 \mbox{(syst)} ~\mbox{ps}.
\end{equation}
The obtained $\Lb$ lifetime is 3.4 standard deviation larger than
the world average computed excluding this result. Therefore, additional measurements
of the $\Lb$ lifetime are required to resolve this ambiguity.

In addition to $\Lb$ lifetime, the CDF collaboration also measured\cite{omb-cdf} the lifetime of $\Xibm$ and $\Ombm$
baryons. The obtained values are
\begin{eqnarray}
\tau(\Xibm) & = & 1.56 ^{+0.27}_{-0.25}\mbox{(stat)} \pm 0.02 \mbox{(syst)} ~\mbox{ps}. \\
\tau(\Ombm) & = & 1.13 ^{+0.53}_{-0.40}\mbox{(stat)} \pm 0.02 \mbox{(syst)} ~\mbox{ps}.
\end{eqnarray}
The precision of these measurements is not sufficient to make any conclusion on a possible variation
of the lifetime of different $B$ baryons, so that more precise measurements at LHC are
required to clarify the lifetime pattern in $B$ baryons.

\subsection{Comparison with theoretical predictions and conclusions}

The results on the lifetime of different $B$ hadrons obtained at the Tevatron provide a very interesting
possibility to verify the theoretical predictions. The most accurate values predicted by the
theoretical models are given for the ratios of $B$-hadron lifetimes relative
to the $\Bd$ lifetime, and namely the comparison of these ratios
is shown in Table \ref{tab2}. The world average experimental results and the theoretical
predictions given in this Table are
taken from Ref.\ \cite{hfag}.
\begin{table}[tbph]
\begin{center}
\caption{
Measured ratios of $B$-hadron lifetimes relative to the $\Bd$ lifetime and ranges predicted
by theory.
}
{\begin{tabular}{@{}|c|c|c|@{}} 
\hline
Lifetime ratio & Measured value & Predicted range \\
\hline
$\tau(\Bp)/\tau(\Bd)$ & $1.079 \pm 0.007$ & $1.04 - 1.08$ \\
$\tau(\Bs)/\tau(\Bd)$ & $0.993 \pm 0.009$ & $0.99 - 1.01$ \\
$\tau(\Lb)/\tau(\Bd)$ & $0.930 \pm 0.020$ & $0.86 - 0.95$ \\
$\tau(\Bcp)/\tau(\Bd)$ & $0.302 \pm 0.020$ & $0.24 - 0.35$ \\
\hline
\end{tabular}
\label{tab2}}
\end{center}
\end{table}

There is a striking agreement between the experiment and the theory for all types of $B$ hadrons.
The experiments confirm the expected pattern of $B$-hadron lifetimes given in Eq.\ (\ref{pattern}).
It can also be noticed that the experimental precision is currently better for all $B$ hadrons,
which opens an excellent opportunity of improving the theoretical computations.
Many Tevatron results were obtained with a small part of the available statistics, so that the precision
of many Tevatron measurements could be significantly improved. However, the lack of
manpower will probably prevent to achieve this improvement. Nevertheless,
it is important to stress that the contribution of the Tevatron results, even without adding more
statistics, is dominant or essential for the precision of all measured $B$-hadron lifetimes.
Thus, the lifetimes of $B$ hadrons
is a very important achievement and a long-lasting legacy of the CDF and D\O\ experiments.

\section{$\Bs - \Bsbar$ mixing}
\label{mixing}

All neutral ground state mesons ($K^0$, $D^0$, $\Bd$, $\Bs$) can change their flavor from particle to antiparticle
during the lifetime. This phenomenon is called an oscillation.
The $\Bd$ oscillations was well established before the experiments at the Tevatron\cite{hfag},
with a precisely measured oscillation frequency $\Delta m_d$.
In the SM, the parameter $\Delta m_q$ of $B_q^0$ meson, where where $q = d,s$,
is proportional to the combination $|V_{tb}^* V_{tq}|^2$ of Cabibbo-Kobayashi-Maskawa (CKM) matrix
elements\cite{pdg-2012}. Since the matrix element $V_{ts}$ is larger than $V_{td}$,
the frequency $\Delta m_s$ is higher, which makes it very difficult to detect. As a result, the $\Bs$
oscillation have not been observed by any previous experiment.
However, $\Delta m_s$ is a crucial parameter for establishing the unitarity relation
of the CKM matrix. Its measurement yields the ratio $|V_{ts}/V_{td}|$, which has a smaller uncertainty than
$|V_{td}|$ alone due to the cancellation of certain theory uncertainties and provides a stringent
constraint on the unitarity triangle and the source of CP violation in the SM\cite{pdg-2012,ckmfitter,utfit,utfit-1}.

The measurement of the oscillation frequency is very complicated, since it requires the identification
of both initial and final state of the $\Bs$ meson and the detection
of the time evolution of oscillated events. The D\O\ collaboration used for this purpose\cite{dms-d0}
the semileptonic $\Bs \to \mu+ D_s^- X$ decays with $D_s^- \to \phi \pi^-$. The analysed statistics
corresponds to the integrated luminosity of 1 fb$^{-1}$. They collected $26700 \pm 556$ candidates.

The final flavor of $B_s$ meson in this decay mode is determined by the charge of the muon.
The initial flavor tagging is more involved. Its quality is described by the tagging power $P$, which
is defined as $P = \varepsilon (2 f_R - 1)^2$, where $\varepsilon$ is the efficiency to select the tagged events
and $f_R$ is the fraction of events with correct identification of the initial flavor relative to the
total number of tagged events. The tagging power $P$ multiplied by the total number of selected events
corresponds to the effective statistics used in the measurement of oscillation frequency.

The D\O\ collaboration developed a sophisticated initial flavor tagging technique\cite{ft-d0}. It is based
on the measurement of the flavor of the complementary $B$ hadron in the pair production $\Bs \bar B$ or $\Bsbar b$.
Its flavour is correlated with the initial flavour of $\Bs$ meson. This technique is called
the opposite-side flavor tagging. The D\O\ collaboration achieved the tagging power
\begin{equation}
P = (2.48 \pm 0.21 \mbox{(stat)} ^{+0.08}_{-0.06} \mbox{(syst)} )\times 10^{-2}.
\end{equation}

\begin{figure}[tpbh]
\begin{center}
\centerline{
\psfig{file=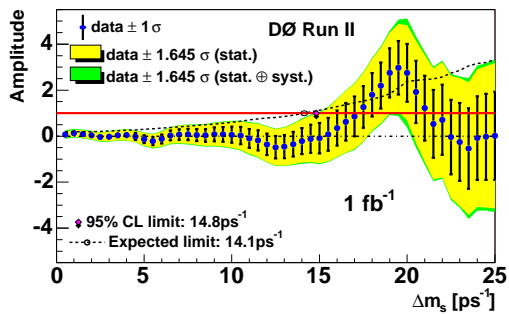,width=0.60\textwidth}
}
\vspace*{8pt}
\caption{
$\Bs$ oscillation amplitude as a function of oscillation
frequency, $\Delta m_s$, taken from Ref. \cite{dms-d0}. The solid line shows the $A = 1$ axis for
reference. The dashed line shows the expected limit including
both statistical and systematic uncertainties.
%Value of $−\Delta \log \cal{L}$ as a function of $\Delta m_s$. Star symbols
%do not include systematic uncertainties, and the shaded band
%represents the envelope of all $\log \cal{L}$ scan curves due to different
%systematic uncertainties.
\label{dms-amp-d0}}
\end{center}
\end{figure}

The flavor-tagged events were used in an unbinned fitting procedure.
The likelihood, $\cal L$, for an event to arise from a specific
source of the signal or background was determined event-by-event. This likelihood depends on the measured
decay length, its uncertainty , the invariant mass of the candidate
$m(K^+K^- \pi^-)$, the predicted value of $f_R$, and the purity of the signal selection.
The resulting amplitude of $\Bs$ oscillation as a function of $\Delta m_s$ is
shown in Fig.\ \ref{dms-amp-d0}.
%The change of this likelihood relative to an arbitrary level, $-\Delta \log \cal{L}$ as a function of $\Delta m_s$
%is shown in Fig.\ \ref{dms-ll-d0}. The variation of $-\Delta \log \cal{L}$ by one unit corresponds
%to the region of one standard deviation in the value of $\Delta m_s$.
The amplitude $A = 1$ corresponds to the $\Bs$ oscillation, and the amplitude $A=0$ corresponds
to the absence of oscillation.
Using the selected statistics and the initial flavor tagging, the D\O\ collaboration
determined the 90\% confidence level interval of the $\Bs$ oscillation frequency
\begin{equation}
17 < \Delta m_s < 21 ~\mbox{ps}^{-1}~\mbox{(90\% C.L.).}
\end{equation} At the time of publication it was the first double-sided bound on this quantity.

\begin{figure}[tpbh]
\begin{center}
\centerline{
\psfig{file=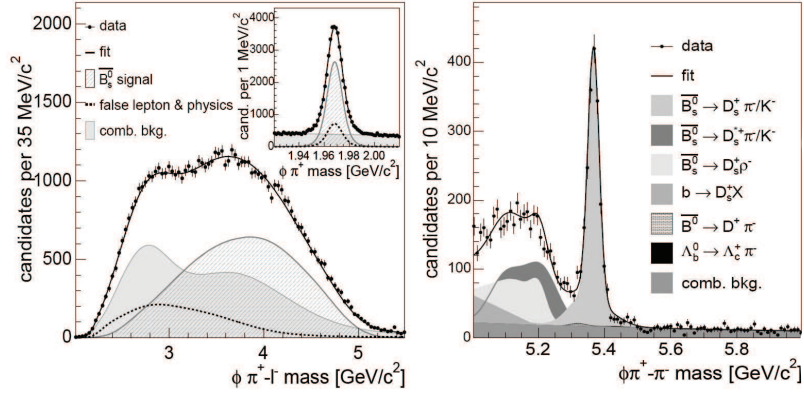,width=0.90\textwidth}
}
\vspace*{8pt}
\caption{
(Left panel) The invariant mass distributions for the $D_s^+ \to \phi \pi^+$ candidates [inset]
and the $l^- D_s^+ (\phi \pi^+)$ pairs. The contribution labelled ``false lepton \& physics'' refers
to backgrounds from hadrons mimicking the lepton signature combined
with real $D_s$ mesons and physics backgrounds such as
$\Bd \to D_s^+ D^-$, $D_s^+ \to \phi \pi^+$, $D^- \to l^- X$.
(Right panel) The invariant mass distribution for $\Bsbar \to D_s^+ (\phi \pi^+) \pi^-$
decays including the contributions from $\Bsbar \to D_s^{*+} \pi^-$ and
$\Bsbar \to D_s^+ \rho^-$. In this
panel, signal contributions are drawn added on top of the combinatorial background.
The plots are taken from Ref. \cite{dms-cdf}.
\label{dms-mass}}
\end{center}
\end{figure}

The CDF collaboration searched for the $\Bs$ oscillation in both semileptonic and hadronic final state.\cite{dms-cdf}
They used the statistics corresponding to 1 fb$^{-1}$ of $p \bar p$ collisions. The hadronic decays
include $\Bs \to D_s^- \pi^+$ and $\Bs \to D_s^- \pi^+ \pi^- \pi^+$. The semileptonic decay mode
used in the analysis is $\Bs \to l^+ D_s^- X$ ($l= \mu, e$).  The $D_s^-$ meson decays to three
different final states $D_s^- \to \phi \pi^-$, $D_s^- \to K^{*0} K^-$, and $D_s^- \to \pi^+ \pi^- \pi^-$ was used.
In addition, the partially reconstructed hadronic decays with one or two photons missing were included in the analysis.
These decays are $\Bs \to D_s^{*-} \pi$, $D_s^{*-} \to D_s^- \pi$ and $\Bs \to D_s^- \rho^+$, $\rho^+ \to \pi^+ \pi^0$
with $D_s^- \to \phi \pi^-$. Their addition significantly increases the statistics, which is essential
for this analysis. In total, 5600 fully reconstructed hadronic $\Bs$ decays, 3100 partially reconstructed
hadronic $\Bs$ decays and 61500 partially reconstructed semileptonic $\Bs$ decays were selected.
Figure\ \ref{dms-mass} shows the invariant
mass distribution of $\Bs \to \mu^+ D_s^- X$ (left plot) and $\Bs \to D_s^- \pi^+$ (right plot) candidates.
This plots show the contribution of different channels in the selected event samples.
They also demonstrate an excellent understanding of the sample composition achieved in this analysis.

\begin{figure}[tpbh]
\begin{center}
\centerline{
\psfig{file=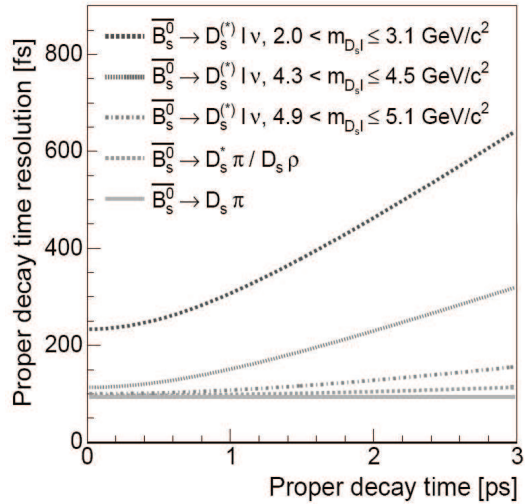,width=0.60\textwidth}
}
\vspace*{8pt}
\caption{
The average proper decay time resolution for $\Bs$ decays as a function of proper
decay time from Ref. \cite{dms-cdf}.
\label{dms-resol}}
\end{center}
\end{figure}

The fully and partially reconstructed hadronic decays give an important advantage to the CDF analysis.
The precision of the proper decay time reconstruction is much better for these decays than for the semileptonic
decays. The comparison of this precision for different decay types is shown in Fig.\ \ref{dms-resol}
taken from Ref.\ \cite{dms-cdf}.
This precision is essential for the measurement of the $\Bs$
oscillation. The $\Bs$ meson changes flavor with high frequency, and namely this change needs to be
detected. The value of $\Delta m_s = 17.5$ ps$^{-1}$ corresponds to the
period of oscillation in the proper decay time $\tau = 2 \pi / \Delta m_s \simeq 360 $ fs. The precision
on this quantity should be at least 4 times better to measure the oscillation reliably. That is why
the trigger on displaced tracks, allowing selection of the hadronic $\Bs$ decays, played a crucial role
in the successful detection of the $\Bs$ oscillation by the CDF collaboration.

Another important feature of this analysis is the application of both opposite-side and same-side
flavor tagging. The $\Bs$ meson is often accompanied
by the charged kaon, produced during hadronisation of the initial $b$ quark.
Its charge corresponds to the initial flavor of the $\Bs$ meson. Thus, by selecting
this additional kaon and measuring its charge it is possible to determine the initial $\Bs$ flavor.
This technique is called the same-side flavor tagging. To implement it, the kaon needs to be identified
among all other charged tracks. In the CDF analysis, the measurement of energy loss
in the tracking detector and time-of-ﬂight information are used to identify the kaons.
The tagging power $P= 3.7\% (4.8\%)$ is achieved for the same-side tagging in hadronic (semi-leptonic) decay sample.
The fractional uncertainty on $P$ is approximately 25\%.
The opposite-side tagging power in the CDF measurement is $P= 1.8 \pm 0.1\%$.

\begin{figure}[tpbh]
\begin{center}
\centerline{
\psfig{file=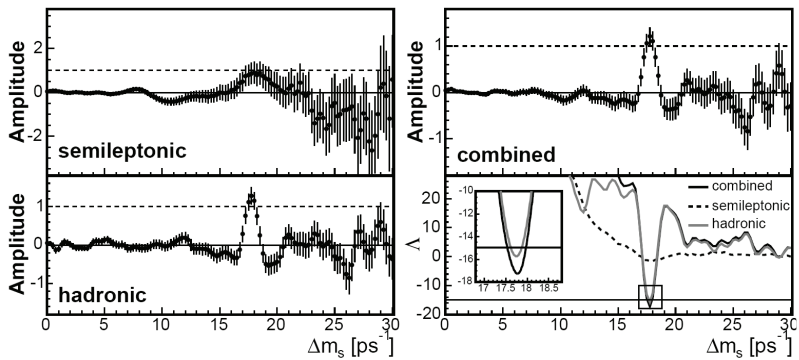,width=0.90\textwidth}
}
\vspace*{8pt}
\caption{
The measured amplitude values and uncertainties versus the $\Bs - \Bsbar$ oscillation frequency $\Delta m_s$
taken from Ref. \cite{dms-cdf}.
(Upper Left) Semileptonic
decays only. (Lower Left) Hadronic decays only.
(Upper Right) All decay modes combined. (Lower Right)
The logarithm
of the ratio of likelihoods for amplitude equal to one and amplitude equal to zero,
$\Lambda = \log[{\cal L}^{A=0}/{\cal L}^{A=1}(\Delta m_s)]$, versus the
oscillation frequency. The horizontal line indicates the value $\Lambda = −15$ that corresponds
to a probability of $5.7 × 10^{−7}$ $(5\sigma)$ in
the case of randomly tagged data.
\label{dms-scan}}
\end{center}
\end{figure}

The CDF collaboration uses an unbinned maximum likelihood fit to search for Bs oscillations.
The likelihood combines mass, decay time, decay-time resolution, and flavor tagging information for each candidate,
and includes terms for signal and each type of background.
The resulting amplitude of $\Bs$ oscillation as a function of $\Delta m_s$ is
shown in Fig.\ \ref{dms-scan}. It can be seen that the amplitude increases to ${\cal{A}} = 1.20 \pm 0.20$ in the
region around $\Delta m_s = 17.75$ ps$^{-1}$. For all other values of $\Delta m_s$ it is consistent
with zero. Using the available statistics and the developed analysis the CDF collaboration
was able to report the discovery of the $\Bs$ oscillation. The measured oscillation frequency
was found to be
\begin{equation}
\Delta m_s = 17.77 \pm 0.10 \mbox{(stat)} \pm 0.07 \mbox{(syst)} ~\mbox(ps)^{-1}.
\end{equation}

\begin{figure}[tpbh]
\begin{center}
\centerline{
\psfig{file=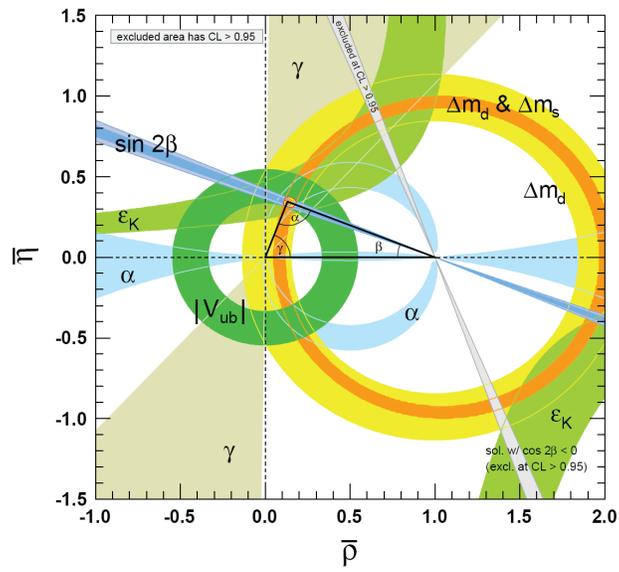,width=0.70\textwidth}
}
\vspace*{8pt}
\caption{
Constraints on the parameters of CKM matrix taken from Ref.\ \cite{pdg-2012}.
The shaded areas have 95\% confidence level.
\label{ckm-pdg}}
\end{center}
\end{figure}

Figure\ \ref{ckm-pdg} taken from Ref.\ \cite{pdg-2012}
shows the impact of the $\Delta m_s$ measurement on the unitarity test of the CKM matrix.
It can be seen that this result provides one of the most strong constraints and is essential
for verifying the unitarity of the CKM matrix. Thus, the measurement of the oscillation frequency
at the Tevatron is one of the most important achievement of its $B$ physics program.

\section{Decays of $B$ hadrons}
\label{rare}

The decays of $B$ hadrons are very numerous and offer rich possibilities of various studies.
It is impossible to cover all results in one review due to the size limitations.
We concentrate here on the most interesting results obtained at the Tevatron.

Probably the most important measurement in this research area is the search for the rare
decays $\Bd \to \mu^+ \mu^-$ and $\Bs \to \mu^+ \mu^-$. These flavor-changing neutral current (FCNC)
decays are forbidden in the SM at the tree level.
Therefore, the SM predicts a very low value for the branching fractions of both
$B^0 \to \mu^+ \mu^-$ and $B_s^0 \to \mu^+ \mu^-$ decays. The most recent SM prediction
for these fractions is \cite{gamiz,buras,buras-1}
\begin{eqnarray}
\mbox{Br}(B_s^0 \to \mu^+ \mu^-) & = & (3.23 \pm 0.27) \times 10^{-9}, \nonumber \\
\mbox{Br}(B^0   \to \mu^+ \mu^-) & = & (1.07 \pm 0.10) \times 10^{-10}.
\end{eqnarray}
The contribution of new physics beyond the SM can significantly modify these
values.\cite{hamzaoui,choudhury,babu,buras1}
Thus, these rare decays can provide important constraints on various new physics models.

The D\O\ collaboration searched for these decays using 6.1 fb$^{-1}$ of available statistics.\cite{bs-mumu-d0}
A multivariate neural network analysis was used to separate the possible signal from the background.
Using this statistics, the following 95\% confidence level upper limit on the
branching fraction $\Bs \to \mu^+ \mu^-$ was obtained:
\begin{equation}
\mbox{Br}(\Bs  \to \mu^+ \mu^-)  <  5.1 \times 10^{-9}~ \mbox{at 95\% C.L.}
\end{equation}
The insufficient resolution of D\O\ tracking system did not allow to separate the possible contribution
of $\Bd \to \mu^+ \mu^-$ decay. Therefore, the above result was obtained assuming the SM value
of $\Bd \to \mu^+ \mu^-$ branching fraction.

\begin{figure}[tpbh]
\begin{center}
\centerline{
\psfig{file=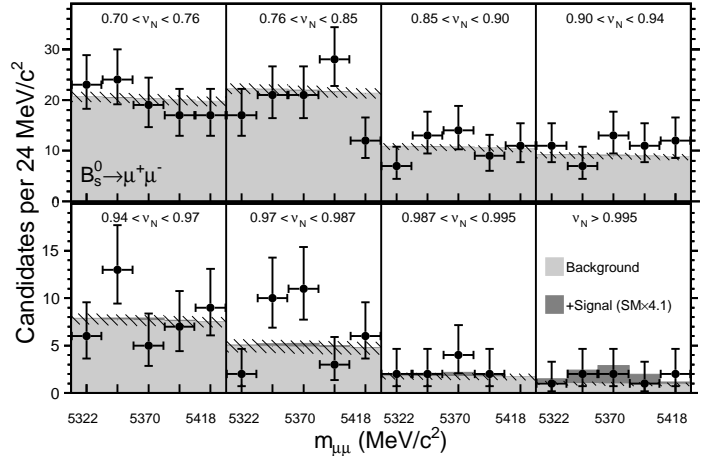,width=0.80\textwidth}
}
\vspace*{8pt}
\caption{
For the $B_s$ mass region, the observed number of events (points) is compared to the
total expected background (light grey) and its uncertainty (hatched) for different values of $\nu_N$,
taken from Ref. \cite{mumu-cdf-2}.
The hashed area represents the systematic uncertainty on the mean expected background while the error
bars on the points represent the associated poisson uncertainty.
Also shown is the expected contribution from $\Bs \to \mu^+ \mu^-$ events (dark gray) using a
branching fraction that corresponds to the central value from the fit to the data,
which is 4.1 times the expected SM value.
\label{bs-mumu-cdf}}
\end{center}
\end{figure}

The CDF collaboration presented in summer 2011 the analysis \cite{mumu-cdf-1}
with 7 fb$^{-1}$ featuring an accumulation of signal-like events in the $B_s^0$ mass region
with $\sim 2.5 \sigma$ deviation from the background-only hypothesis.
The latest CDF analysis,\cite{mumu-cdf-2} which was still unpublished at the time of preparing this review,
includes the full Run2 statistics corresponding
to 9.6 fb$^{-1}$. Given the increased interest to the previous result,
the analysis of the remaining statistics is kept the same. The separation between the signal and
background in this analysis is achieved using the neural network. Figure \ref{bs-mumu-cdf} shows the observed
and expected number of events in the $B_s^0 \to \mu^+ \mu^-$ search
for the different values of the neural network output variable $\nu_N$.
There is an excess of the signal-like events for $\nu_N > 0.97$,
while the agreement between the observed and expected number of events is very good for
the background-dominated region $\nu_N < 0.97$. The $p$-value of the SM signal plus background hypothesis for
$\nu_n > 0.97$ is 7\%. The excess of events in the $0.97 < \nu_N < 0.987$ bin is not increased with the
addition of the new statistics and is consistent with the statistical fluctuation. The $p$-value
of the SM signal plus background hypothesis for two largest $\nu_N$ bins is 22.4\%, while the $p$-value of
background only hypothesis is 2.1\%. Thus, while still not conclusive, the experiment becomes
sensitive to the SM contribution of $B_s^0 \to \mu^+ \mu^-$ decay and shows a good agreement with
the SM expectation.

The results obtained by the CDF collaboration with 9.6 fb$^{-1}$ are:
\begin{eqnarray}
\mbox{Br}(B_s^0 \to \mu^+ \mu^-) & = & (1.3^{+0.9}_{-0.7}) \times 10^{-8}, \nonumber \\
\mbox{Br}(B^0   \to \mu^+ \mu^-) & < & 4.6 \times 10^{-9}~ (3.8 \times 10^{-9})~ \mbox{at 95\% (90\%) C.L.}
\end{eqnarray}
The CDF collaboration also reports the first double sided limit on $\mbox{Br}(B_s^0 \to \mu^+ \mu^-)$:
\begin{eqnarray}
0.8 \times 10^{-9}  < & Br(B_s^0 \to \mu^+ \mu^-) & <  3.4 \times 10^{-8} ~ \mbox{at 95\% C.L.}, \nonumber \\
2.2 \times 10^{-9}  < & Br(B_s^0 \to \mu^+ \mu^-) & <  3.0 \times 10^{-8} ~ \mbox{at 90\% C.L.}
\end{eqnarray}
%These results are consistent with other searches of these rare decays.

The search of these rare decays continued at LHC and the LHCb collaboration reported recently the evidence
of the $\Bs \to \mu^+ \mu^-$ decay with the branching fraction consistent with the SM
expectation:\cite{bs-mumu-lhcb}
\begin{equation}
Br(\Bs  \to \mu^+ \mu^-)  =  (3.2 ^{+1.5}_{-1.2}) \times 10^{-9}.
\end{equation}
The results of D\O\ and CDF are consistent with the value obtained
by the LHCb collaboration. All these results are consistent with the SM prediction, providing a strong constraint on the
contribution of the new physics processes.

The CDF collaboration performed an extensive study of the decays mediated by the FCNC transition
$b \to s \mu^+ \mu^-$ with different initial and final hadrons. The analysis is based on the statistics
corresponding to 9.6 fb$^{-1}$ of $p \bar p$ collisions. This result\cite{b-to-s-cdf}
was still unpublished at the time of preparing this report. In each case the branching fraction of the decay
$H_b \to h \mu^+ \mu^-$ is normalised to the well identified decay $H_b \to J/\psi h$ with $J/\psi \to \mu^+ \mu^-$.
Such a normalisation significantly reduces the systematic uncertainty of the measurements.
The following results in different decay modes are obtained:
\begin{eqnarray}
\mbox{Br}(B^+ \to K^+ \mu^+ \mu^-) & = & [0.45 \pm 0.03 \mbox{(stat)} \pm 0.02 \mbox{(syst)}]\times 10^{-6}, \\
\mbox{Br}(B^+ \to K^{*+} \mu^+ \mu^-) & = & [0.89 \pm 0.25 \mbox{(stat)} \pm 0.09 \mbox{(syst)}]\times 10^{-6}, \\
\mbox{Br}(\Bd \to K^0 \mu^+ \mu^-) & = & [0.33 \pm 0.08 \mbox{(stat)} \pm 0.03 \mbox{(syst)}]\times 10^{-6}, \\
\mbox{Br}(\Bd \to K^{*0} \mu^+ \mu^-) & = & [1.14 \pm 0.09 \mbox{(stat)} \pm 0.06 \mbox{(syst)}]\times 10^{-6}, \\
\mbox{Br}(\Bs \to \phi \mu^+ \mu^-) & = & [1.17 \pm 0.18 \mbox{(stat)} \pm 0.37 \mbox{(syst)}]\times 10^{-6}, \\
\mbox{Br}(\Lambda_b \to \Lambda \mu^+ \mu^-) & = & [1.95 \pm 0.34 \mbox{(stat)} \pm 0.61 \mbox{(syst)}]\times 10^{-6}.
\end{eqnarray}

Another interesting study performed at the Tevatron is the measurement of the branching fraction of the
decay $B_s \to D_s^{(*)+} D_s^{(*)-}$. This decay is not rare, but it is expected that the final state is
mainly $CP$-even and may saturate the total decay width of $\Bs$ meson under certain theoretical
assumptions.\cite{alexan}

The D\O\ collaboration made an inclusive measurement of this branching fraction using 1.3 fb$^{-1}$ of
available statistics.\cite{bs-dsds-d0} The decay
$B_s \to D_s^{(*)+} D_s^{(*)-}$ was selected using the inclusive muon trigger. One $D_s$ meson
was partially reconstructed in the $D_s \to \mu \nu \phi $ decay mode. Another $D_s$ meson
was selected in $D_s \to \phi \pi$ decay mode.
No attempt was made to distinguish $D_s$ and $D_s^*$ states.
The resulting branching fraction is
\begin{equation}
\mbox{Br}(B_s \to D_s^{(*)+} D_s^{(*)-}) = 0.039 ^{+0.019}_{-0.017} \mbox{(stat)} ^{+0.016}_{-0.015} \mbox{(syst)}.
\end{equation}

\begin{figure}[tpbh]
\begin{center}
\centerline{
\psfig{file=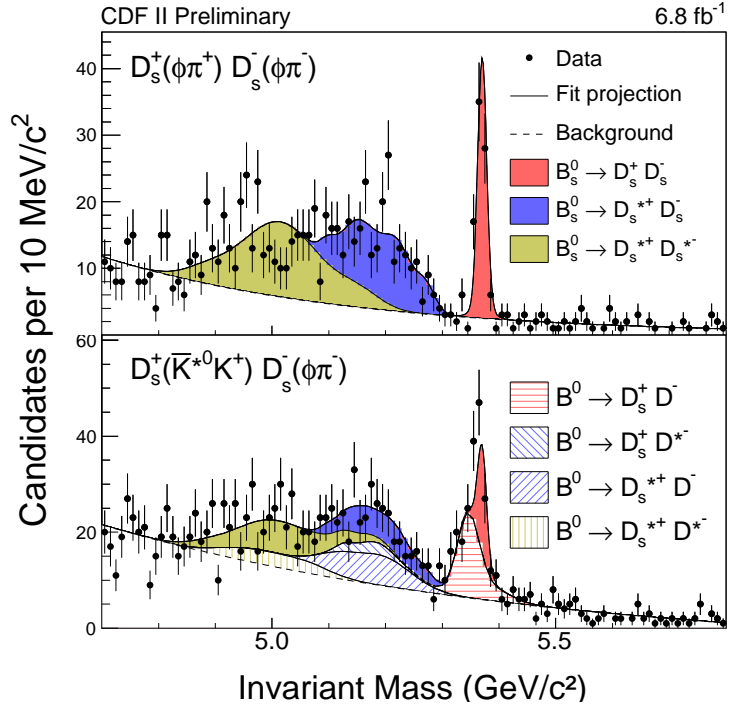,width=0.80\textwidth}
}
\vspace*{8pt}
\caption{
Invariant mass distribution of $B_s^0 \to D_s^{+}(\phi \pi^+) D_s^{-} (\phi \pi^-)$
and $B_s^0 \to D_s^{+}(K^{*0} K^+) D_s^{-} (\phi \pi^-)$. The plot is taken from Ref. \cite{bs-dsds-cdf}.
\label{bs-dsds-mass-cdf}}
\end{center}
\end{figure}

The CDF collaboration performed an analysis using 6.8 fb$^{-1}$ of statistics.\cite{bs-dsds-cdf}
They exclusively reconstructed all decay modes using the hadronic decays $D_s \to \phi \pi$ or $D_s \to K^* K$.
The resulting invariant mass distribution is presented in Fig.~\ref{bs-dsds-mass-cdf}.
In total 750 signal events in these decay modes are reconstructed.
Using this statistics, the following result is obtained
\begin{eqnarray}
Br(B_s^0 \to D_s^{+} D_s^{-}) & = & (0.49 \pm 0.06 \mbox{(stat)} \pm 0.05 \mbox{(syst)} \pm 0.08\mbox{(norm)}) \%,  \\
Br(B_s^0 \to D_s^{*\pm} D_s^{\mp}) & = & (1.13 \pm 0.12 \mbox{(stat)} \pm 0.09 \mbox{(syst)} \pm 0.19 \mbox{(norm)} )\%, \\
Br(B_s^0 \to D_s^{*+} D_s^{*-}) & = & (1.75 \pm 0.19 \mbox{(stat)} \pm 0.17 \mbox{(syst)} \pm 0.29 \mbox{(norm)} )\%.
\end{eqnarray}
The total branching fraction of these decay modes is found to be
\begin{equation}
Br(B_s^0 \to D_s^{(*)+} D_s^{(*)-}) =  (3.38 \pm 0.25 \mbox{(stat)} \pm 0.30 \mbox{(syst)} \pm 0.56 \mbox{(norm)} )\%.
\end{equation}
The results obtained by the CDF and D\O\ collaborations are consistent.

\section{Study of $CP$ asymmetry with $B$ decays}
\label{cpv}

An important part of $B$ physics research at the Tevatron was devoted to the
measurement of the $CP$ asymmetry.
Among other reasons, the interest to this phenomenon is explained by the fact that
the magnitude of the $CP$ asymmetry included in the SM is not sufficient to describe the observed abundance
of matter in our universe \cite{huet}, which implies that some additional sources
of $CP$ asymmetry should exist.
They could reveal themselves by deviating the observed $CP$ asymmetry from the SM prediction.
While the $CP$ asymmetry in the decays of $\Bd$ and $\Bp$ mesons was extensively
studied at $B$ factories, the experiments at the Tevatron offer a possibility to study the $CP$ violation
in the decays of $\Bs$ mesons. Until the start of the LHC era, it was the only place where such measurements
were possible.

\subsection{$CP$ asymmetry in the decay $\Bs \to J/\psi \phi$}

One of the most promising channels to search for the new sources of $CP$ asymmetry is the
decay $B_s^0 \to J/\psi \phi$. The $CP$ asymmetry in this decay is described by the phase $\phi^{J/\psi \phi}$.
Within the SM, this phase is related with the angle $\beta_s$ of the $(bs)$ unitarity triangle
and is predicted to be very small \cite{lenz}:
\begin{equation}
\phi_s^{J/\psi \phi}(SM) = -2 \beta_s = -0.036 \pm 0.002.
\end{equation}
This phase can be significantly modified by the new physics contribution and this deviation from the SM
can be detected experimentally.

\begin{figure}[tpbh]
\begin{center}
\centerline{
\psfig{figure=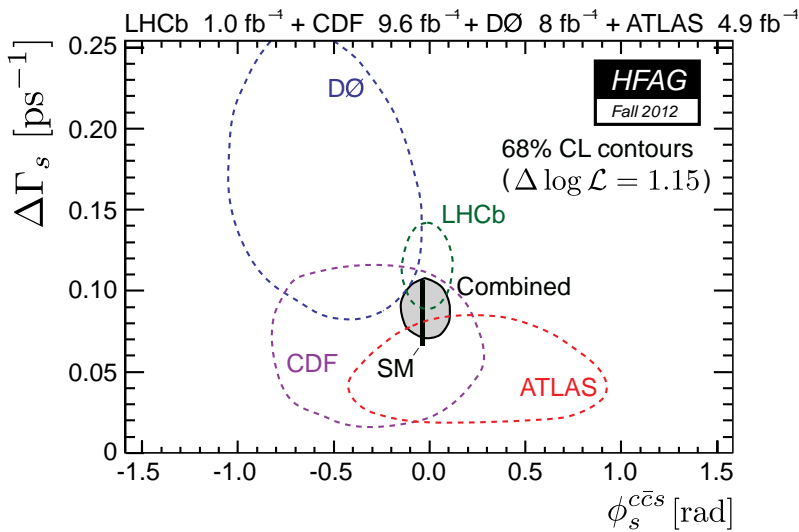,height=7cm}
}
\vspace*{8pt}
\caption{
Constraints of all measurements of CP violation in $\Bs \to J/\psi \phi$ in the
$\phi_s^{J/\psi \phi}$--$\Delta \Gamma_s$  plane taken from Ref. \cite{hfag}.
\label{phis-comb}}
\end{center}
\end{figure}

Both CDF and D\O\ experiments report their final results on the
$CP$ asymmetry in the $B_s^0 \to J/\psi \phi$ decay with the full statistics.
The CDF collaboration reconstructs \cite{cdf-jpsi} about 11000 such decays using the integrated
luminosity 9.6 fb$^{-1}$.
%The new analysis \cite{cdf-jpsi} is similar to
%the previous measurement with a part of the statistics \cite{cdf-jpsi-1}.
The result of this analysis
is presented in Fig. \ref{phis-comb} as the confidence regions in $\phi^{J/\psi \phi} - \Delta \Gamma_s$ plane.
It can be seen that the obtained confidence region is consistent with the SM prediction within 1$\sigma$.
The obtained confidence regions for the quantity $\beta_s^{J/\psi \phi} \equiv -\phi_s^{J/\psi \phi}/2$ is
\begin{eqnarray}
\beta_s^{J/\psi \phi} & \in & [-\pi/2, -1.51] \cup [-0.06, 0.30] \cup [1.26, \pi/2] ~ \mbox{at 68\% C.L.} \nonumber \\
\beta_s^{J/\psi \phi} & \in & [-\pi/2, -1.36] \cup [-0.21, 0.53] \cup [1.04, \pi/2] ~ \mbox{at 95\% C.L.}
\end{eqnarray}

%\begin{figure}[tpbh]
%\begin{center}
%\centerline{
%\psfig{figure=jpsiphi-cdf.eps,height=5.7cm}
%\psfig{figure=jpsiphi-d0.eps,height=3.9cm}
%}
%\vspace*{8pt}
%\caption{
%Confidence regions in $\phi^{J/\psi \phi} - \Delta \Gamma_s$ plane.
%Left plot: the measurement of CDF collaboration.
%The solid (blue) and dot-dashed (red) contours show the 68\% and
%95\% confidence regions, respectively. The shaded (green) band is the theoretical
%prediction of mixing-induced $CP$ asymmetry.
%Right plot: the measurement of D\O\ collaboration. Two-dimensional 68\%, 90\% and 95\%
%C.L. contours including systematic uncertainties are shown.
%The SM expectation is indicated as a point with an error.
%\label{jpsiphi}}
%\end{center}
%\end{figure}

A similar analysis of $B_s^0 \to J/\psi \phi$ decay by the D\O\ collaboration \cite{d0-jpsi}
is based on 6500 signal events
collected using the integrated luminosity 8 fb$^{-1}$. The result of this analysis
is shown in Fig. \ref{phis-comb}. The obtained confidence region is consistent
with the SM prediction, and the $p$-value for the SM point is 29.8\%.
The following values are obtained in this analysis:
\begin{eqnarray}
\tau_s & = & 1.443_{-0.035}^{+0.038}~\mbox{ps}, \nonumber \\
\Delta \Gamma_s & = & 0.163^{+0.065}_{-0.064} ~\mbox{ps}^{-1}, \nonumber \\
\phi_s^{J/\psi \phi} & = & -0.55^{+0.38}_{-0.36}.
\end{eqnarray}

Both results are consistent with each other and with the SM prediction.
The latest result from the LHC experiments ATLAS \cite{jpsi-atlas} and LHCb\cite{jpsi-lhcb}
agrees with the values obtained
at the Tevatron. The precision of the LHCb measurement is particularly good, since
it is an experiment dedicated to the $B$ physics studies.
Using a data sample of 0.37 fb$^{-1}$, they obtained
\begin{eqnarray}
\Gamma_s & = & [0.657 \pm 0.009 ~\mbox{(stat)} \pm 0.008 ~\mbox{(syst)} ]~\mbox{ps}^{-1}, \\
\Delta \Gamma_s & = & [0.123 \pm 0.029 ~\mbox{(stat)} \pm 0.011 ~\mbox{(syst)} ]~\mbox{ps}^{-1}, \\
\phi_s^{J/\psi \phi} & = & 0.15  \pm 0.18 ~\mbox{(stat)} \pm 0.06 ~\mbox{(syst)}.
\end{eqnarray}
The combination of all results on the CP violation in the $\Bs \to J/\psi \phi$ decay
is performed by the Heavy Flavour Averaging Group (HFAG) \cite{hfag}. The corresponding plot is
shown in Fig.\ \ref{phis-comb}. The agreement of all measurements
in this channel is reasonably good. It is one of the excellent examples of continuity
of the $B$ physics program from Tevatron to LHC.
An excellent agrement of all results with the SM prediction reduces the prospects
of detecting the new physics contribution in this channel.

\subsection{$CP$ asymmetry in mixing of neutral $B$ mesons}

Studies of the $CP$ asymmetry in mixing of neutral $B^0_q$ ($q = d,s$) mesons provide
another possibility to search for the deviations from the SM prediction.
%is another
%important direction of the $B$ physics at the Tevatron.
This type of asymmetry
%The CP violation in mixing of neutral $B^0_q$ ($q = d,s$) mesons
is described by the $CP$ violating phase $\phi_q$, which is defined as
\begin{equation}
\phi_q \equiv \arg \left( - \frac {m_q^{12}}{\Gamma_q^{12}} \right).
\end{equation}

The parameters $m_q^{12}$ and $\Gamma_q^{12}$ are the complex non-diagonal elements of
the mass mixing matrix. They are related to the observable quantities $\Delta m_q$ and $\Delta \Gamma_q$ as
\begin{equation}
\Delta M_q = 2 \left| m_q^{12} \right|, \qquad
\Delta \Gamma_q = 2 \left| \Gamma_q^{12} \right| \cos \phi_q.
\end{equation}
 The CP violating phase $\phi_q$ can be extracted from the charge asymmetry
$\aslq$ for ``wrong-charge"
semileptonic $B^0_q$-meson decay induced by oscillations, which  is defined as
\begin{equation}
\aslq = \frac{\Gamma(\bar{B}^0_q(t)\rightarrow l^+ X) -
              \Gamma(    {B}^0_q(t)\rightarrow l^- X)}
             {\Gamma(\bar{B}^0_q(t)\rightarrow l^+ X) +
              \Gamma(    {B}^0_q(t)\rightarrow l^- X)}.
              \label{aslq}
\end{equation}
This quantity is independent of the decay time $t$, and can be expressed as
\begin{equation}
\aslq = \frac{\left| \Gamma_q^{12} \right|}{\left| M_q^{12} \right|} \sin \phi_q =
\frac{\Delta \Gamma_q}{\Delta M_q} \tan \phi_q.
\label{phiq}
\end{equation}
The SM predicts the values of $\asld$ and $\asls$ which are not detectable with the
current experimental precision \cite{lenz}:
\begin{equation}
\asld |_{\rm SM}= -(4.1 \pm 0.6) \times 10^{-4}, \quad \asls |_{\rm SM} = (1.9 \pm 0.3) \times 10^{-5}.
\end{equation}
Additional contributions to $CP$ violation via
loop diagrams appear in some extensions of the SM \cite{Randall,Hewett,Hou,Soni,Soni-1,Buras-2,Buras-3,Buras-4}
and can result in these asymmetries
within experimental reach.

In experimental measurements the muon is much easier to identify than any other lepton, therefore
all experimental results on the semileptonic charge asymmetry are obtained with $l=\mu$ in
Eq. (\ref{aslq}). The D\O\ experiment performed several measurements of the semileptonic $\Bd$ and $\Bs$
charge asymmetry. The polarities of the toroidal and solenoidal magnetic fields of D\O\ detector
were regularly reversed so that the four solenoid-toroid polarity
combinations were exposed to approximately the same integrated luminosity.
This feature is especially important in the measurements of the charge asymmetry,
because the reversal of magnets polarities allows for a cancellation of first order
effects related with the instrumental asymmetry and the reduction of the corresponding systematic uncertainty.

One of D\O\ results\cite{asl-d0,asl-d0-1} consists in measuring the like-sign dimuon charge asymmetry $\aslb$.
This quantity is defined as
\begin{equation}
A_{\rm sl}^b \equiv \frac{N_b^{++}-N_b^{--}}{N_b^{++}+N_b^{--}}.
\end{equation}
Here $N_b^{++}$ and $N_b^{--}$ represent the number of events containing two $b$ hadrons
decaying semileptonically and producing two positive or two negative muons, respectively.
Assuming that this asymmetry is produced by $CP$ violation in the mixing of $\Bd$ and $\Bs$ mesons, it can
be expressed as
\begin{equation}
\aslb = C_d \asld + C_s \asls,
\end{equation}
where the coefficients $C_d$ and $C_s$ depend on the mean mixing probabilities $\chi_d$ and $\chi_s$
and the production rates of $\Bd$ and $\Bs$ mesons. Using the integrated luminosity of 9.1 fb$^{-1}$, the
D\O\ experiment obtained
\begin{equation}
\aslb = (-0.787 \pm 0.172 \mbox{(stat)} \pm 0.093 \mbox{(syst)}) \%.
\end{equation}
This result differs by 3.9 standard deviation from the SM prediction. From the study of the impact parameter
dependence of the asymmetry, the D\O\ experiment extracted the separate values of $\asld$ and $\asls$
\begin{eqnarray}
\asld & = & (-0.12 \pm 0.52) \%, \nonumber \\
\asls & = & (-1.81 \pm 1.06) \%.
\end{eqnarray}
The correlation $\rho_{ds}$ between these two quantities is
\begin{equation}
\rho_{ds} = -0.799.
\end{equation}
The precision of these quantities is comparable with the available world average measurements.

\begin{figure}
\begin{center}
\centerline{
\psfig{figure=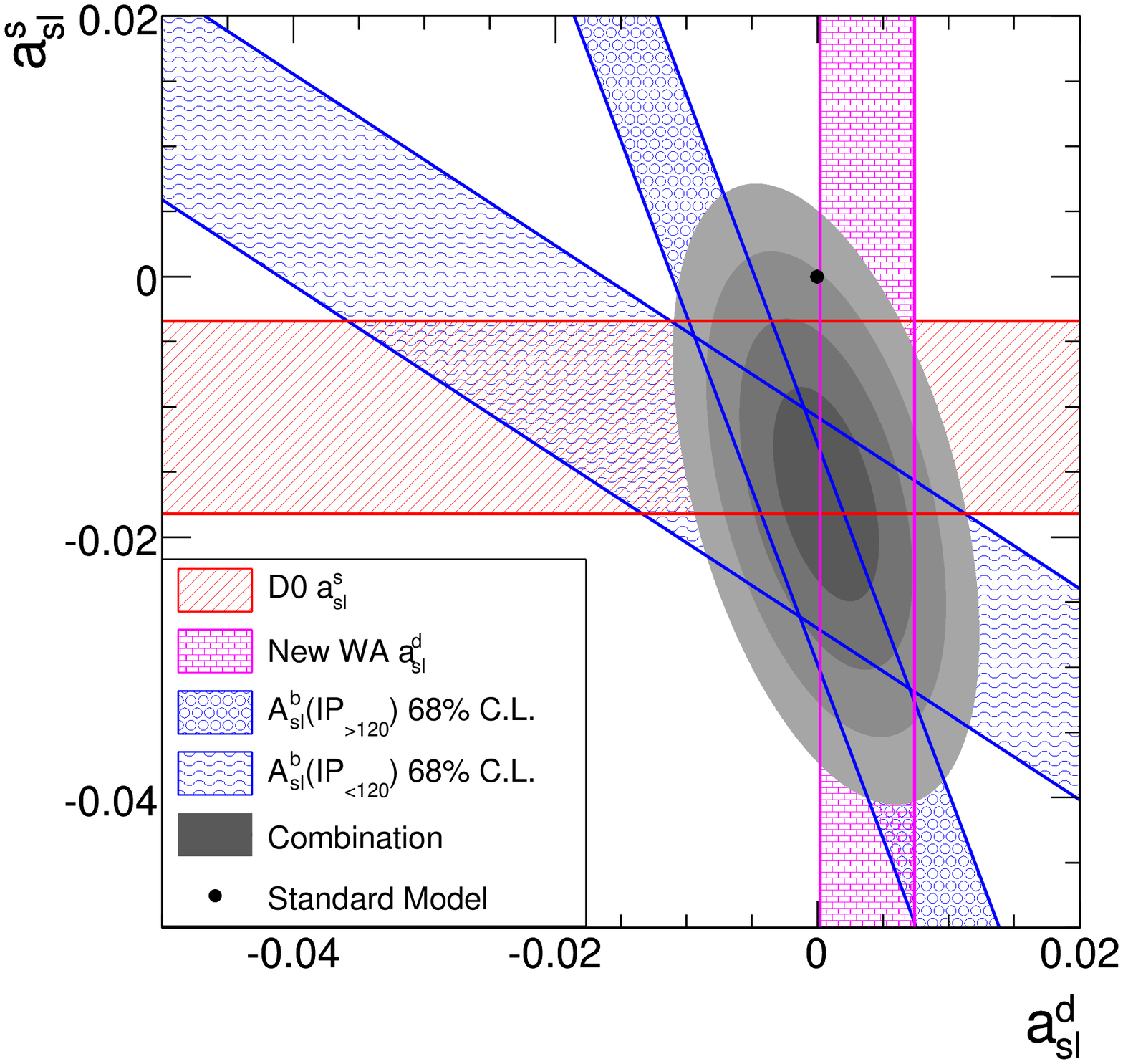,height=6.0cm}
\hspace*{0.3cm}
\psfig{figure=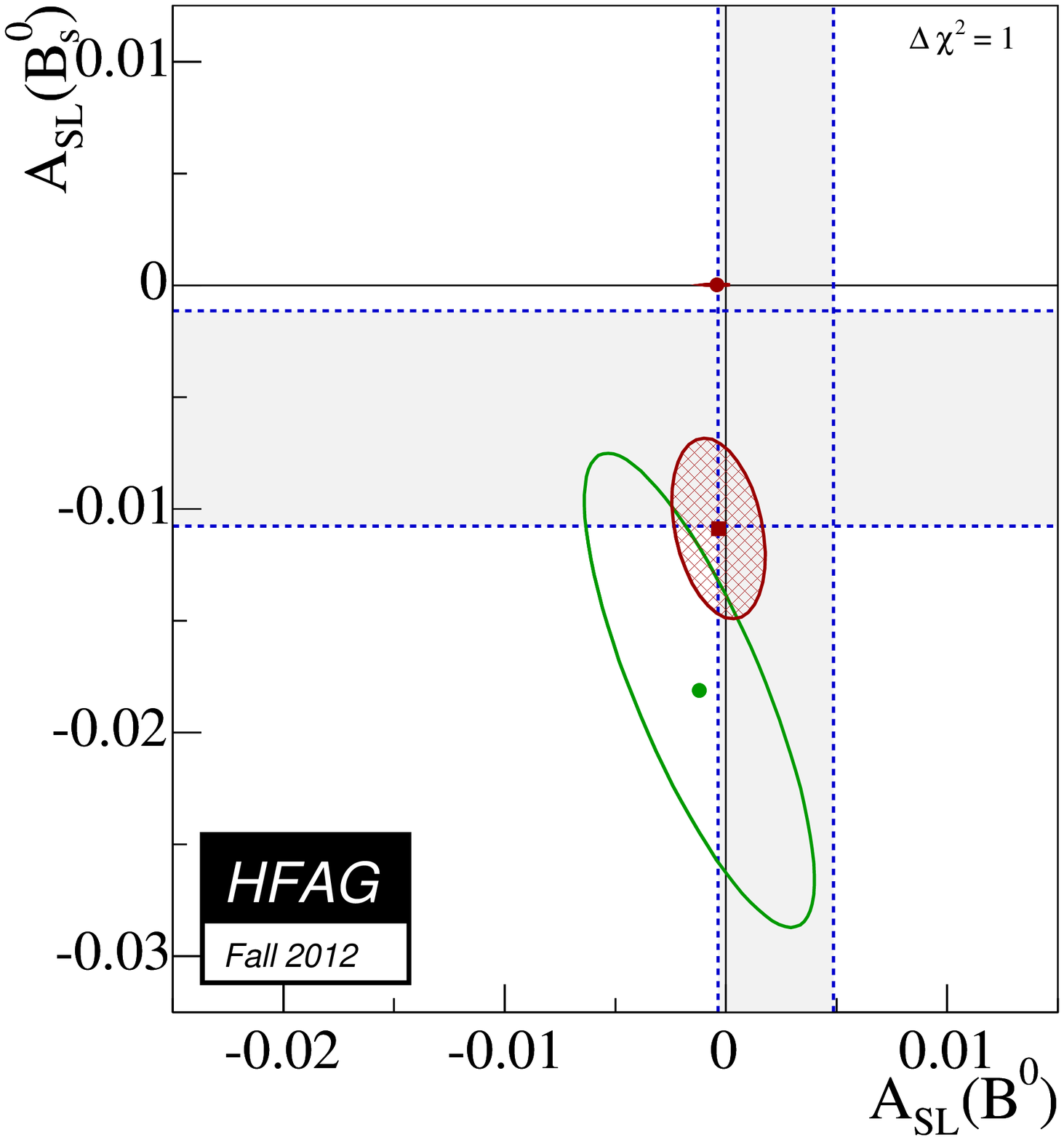,height=7.0cm}
}
\vspace*{8pt}
\caption{
Left plot taken from Ref. \cite{asld-d0}: Combination of measurements of $\asld$ (D0 \cite{asld-d0}
and existing world-average from B factories \cite{pdg-2012}),
$\asls$ (D0 \cite{asls-d0}), and the
two impact-parameter-binned constraints from the same-charge dimuon asymmetry $\aslb$
(D0 \cite{asl-d0}). The bands represent the
$\pm 1$ standard deviation uncertainties on each measurement.
The ellipses represent the 1, 2, 3, and 4 standard deviation two-dimensional
confidence level regions of the combination.
Right plot: Combination of all measurements of $\asld$ and $\asls$ taken from Ref. \cite{hfag}.
the vertical band is the average of the pure $\Bd$ measurements performed at CLEO, BABAR, Belle and D0,
the horizontal band is the average of the pure $\Bs$ measurements performed at D0 and LHCb with semileptonic Bs decays,
the green ellipse is the D0 measurement with same-sign dileptons,
and the read ellipse is the result of the two-dimensional averaging.
The red point close to (0,0) is the SM prediction \cite{lenz} with errors bars multiplied by 10.
\label{asl-d0}}
\end{center}
\end{figure}

The D\O\ experiment also performed separate measurements of the asymmetries $\asld$ and $\asls$ using
the semileptonic decays $\Bd \to \mu^+ \nu D^- X$, $\Bd \to \mu^+ \mu D^{*-} X$,\cite{asld-d0}
and $\Bs \to \mu^+ \nu D_s^- X$ \cite{asls-d0}, respectively. They obtained the following values:
\begin{eqnarray}
\asld = (+0.68 \pm 0.45 \mbox{(stat)} \pm 0.14 \mbox{(syst)}) \%, \\
\asls = (-1.08 \pm 0.72 \mbox{(stat)} \pm 0.17 \mbox{(syst)}) \%.
\end{eqnarray}
Left plot in Fig. \ref{asl-d0}, taken from Ref. \cite{asld-d0}, presents the combination of all results of the D\O\ experiment
on the $CP$ asymmetry in mixing of neutral $B$ mesons, plotted
in the $(\asld, \asls)$ plane. It can be seen that the independent D\O\ measurements
agree well between each other. The combined values of $\asld$ and $\asls$ asymmetries,
including the measurement of $\asld$ asymmetry from $B$ factories \cite{hfag}, is
\begin{eqnarray}
\asld & = & (0.07 \pm 0.27) \%, \nonumber \\
\asls & = & (-1.67 \pm 0.54) \%.
\end{eqnarray}
The correlation $\rho_{ds}$ between these two quantities is
\begin{equation}
\rho_{ds} = -0.46.
\end{equation}
The $p$-value of the combination with respect to the SM point is 0.0037, corresponding to
an inconsistency at the 2.9 standard deviation level.

Recently, the LHCb collaboration performed a similar measurement\cite{asls-lhcb}
of the asymmetry $\asls$ using the decays
$\Bs \to \mu^+ \nu D_s^- X$. They obtained
\begin{equation}
\asls = (-0.24 \pm 0.54 \mbox{(stat)} \pm 0.33 \mbox{(syst)}) \%.
\end{equation}
All these results are consistent within 2 standard deviations,
although the LHCb measurement does not confirm the significant
deviation from the SM observed by the D\O\ experiment. The combination of all available results
on $\asld$ and $\asls$ is performed by HFAG and is given in Ref. \cite{hfag}. The obtained result is
\begin{eqnarray}
\asld & = & (0.03 \pm 0.21) \%, \nonumber \\
\asls & = & (-1.09 \pm 0.40) \%.
\end{eqnarray}
It deviates from the SM prediction by 2.4 standard deviations.
Right plot in Fig. \ref{asl-d0} presents the result of this combination.

\subsection{Other studies of $CP$ asymmetry}

The CDF collaboration performed an extensive study\cite{acp-cdf} of $CP$ asymmetry in the two-body decays
of $B$ hadrons to light hadrons. They used 9.3 fb$^{-1}$ of available statistics.
In particular, they studied the direct CP violation in $\Bs \to K^- \pi^+$ decay.
This decay have been proposed as a clean test for the new physics contribution \cite{Gronau,Lipkin}.
The SM predicts an asymmetry $\sim 30\%$ in this decay.
Any discrepancy from this prediction may indicate the contribution from non-SM amplitudes.

\begin{figure}[tpbh]
\begin{center}
\centerline{
\psfig{figure=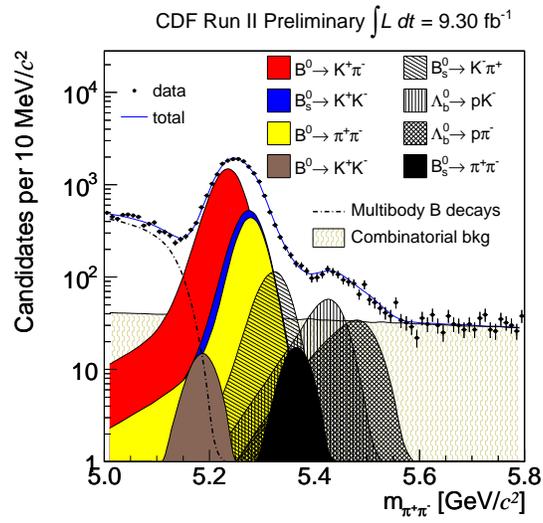,height=7cm}
}
\vspace*{8pt}
\caption{
Mass distribution of reconstructed candidates, $m(\pi \pi)$ taken from Ref. \cite{acp-cdf}.
The charged pion mass is assigned to both particles. The total
projection is overlaid on the data distribution.
\label{pipi-cdf}}
\end{center}
\end{figure}

Since the
particle identification capabilities of the CDF detector are limited, the main task is to disentangle
the contribution of different decays in the invariant mass distribution. Figure\ \ref{pipi-cdf}
shows the result of these efforts. All hadrons are assigned the pion mass,
and the observed $\pi^+ \pi^-$ invariant mass distribution
is presented as the superposition of different two-body decays of $B$ hadrons. The invariant
mass distribution of individual channels is obtained from the simulation.
An unbinned likelihood fit, incorporating kinematic and particle identiﬁcation information is used to
determine the fraction of each individual mode. As a result of this study, the following
values of $CP$ asymmetry in different decay modes were obtained:
\begin{eqnarray}
\frac{\mbox{Br}(\barBd \to K^- \pi^+) - \mbox{Br}(\Bd \to K^+ \pi^-)}
{\mbox{Br}(\barBd \to K^- \pi^+) + \mbox{Br}(\Bd \to K^+ \pi^-)} & = &
-0.083 \pm 0.013 \mbox{(stat)} \pm 0.003 \mbox{(syst)}, \\
\frac{\mbox{Br}(\barBs \to K^+ \pi^-) - \mbox{Br}(\Bd \to K^- \pi^+)}
{\mbox{Br}(\barBs \to K^+ \pi^-) + \mbox{Br}(\Bs \to K^- \pi^+)} & = &
+0.22 \pm 0.07 \mbox{(stat)} \pm 0.02 \mbox{(syst)}, \\
\frac{\mbox{Br}(\Lb \to p \pi^-) - \mbox{Br}(\barLb \to \bar{p} \pi^+)}
{\mbox{Br}(\Lb \to p \pi^-) + \mbox{Br}(\barLb \to \bar{p} \pi^+)} & = &
+0.07 \pm 0.07 \mbox{(stat)} \pm 0.03 \mbox{(syst)}, \\
\frac{\mbox{Br}(\Lb \to p K^-) - \mbox{Br}(\barLb \to \bar{p} K^+)}
{\mbox{Br}(\Lb \to P K^-) + \mbox{Br}(\barLb \to \bar{p} K+)} & = &
-0.09 \pm 0.08 \mbox{(stat)} \pm 0.04 \mbox{(syst)}.
\end{eqnarray}

The D\O\ collaboration performed a measurement\cite{jpsik-d0} of the charge asymmetry in the
$\Bp \to J/\psi K^+ (\pi^+)$ decay using 2.8 fb$^{-1}$ of available statistics.
This asymmetry is defined as
\begin{equation}
A_{CP} =
\frac{\mbox{Br}(\Bm \to J/\psi K^- (\pi^-) - \mbox{Br}(\Bp \to J/\psi K^+ (\pi^+))}
{\mbox{Br}(\Bm \to J/\psi K^- (\pi^-) - \mbox{Br}(\Bp \to J/\psi K^+ (\pi^+))}.
\end{equation}
A non-zero value of $A_{CP}(B^+ \to J/\psi K^+ (\pi^+))$ corresponds
to direct CP violation in this decay. The SM predicts\cite{Hou1} a small
$A_{CP}(B^+ \to J/\psi K^+ \sim 0.003$. Therefore, the observation
of this asymmetry at a higher level would be very interesting.
The results obtained with more than 40000 $\Bp \to J/\psi K^+ (\pi^+)$ decays are:
\begin{eqnarray}
A_{CP}(B^+ \to J/\psi K^+) & = & +0.0075 \pm 0.0061 \mbox{(stat)} \pm 0.0027 \mbox{(syst)}), \\
A_{CP}(B^+ \to J/\psi \pi^+) & = & -0.09 \pm 0.08 \mbox{(stat)} \pm 0.03 \mbox{(syst)}). \\
\end{eqnarray}

\subsection{Summary of the $CP$ asymmetry studies}

In conclusion, an extensive search for the new sources of $CP$ asymmetry is performed
at the Tevatron. Many unique and world best results on the $CP$ asymmetry in the $\Bs$ decays are obtained.
However, no clear indication of the deviation of the $CP$ asymmetry phenomena from the
SM prediction is observed. These studies are continued by the LHC experiment
and the new level of precision in these studies is expected to be achieved.

\section{Conclusions}
\label{end}

A decade of intensive study of $B$ hadrons at the Tevatron produced an impressive list of fascinating results.
Among the milestones achieved at the Tevatron, we can mention the discovery of many
$B$ hadrons and the precise measurement of their properties, including the mass and lifetime of 
$\Bs$ meson and $\Lb$ baryon. The measurement of the oscillation frequency of $\Bs$ meson
provides a crucial constraint on the unitarity of the CKM matrix. The search for
rare decay $\Bs \to \mu^+ \mu^-$, although not attaining the expected SM level, nevertheless imposes
an important constraint on the possible extensions of the SM. Many studies of the $CP$ violation in the
decays and mixing of the $\Bs$ meson are unique or world best. There is an indication of a
possible deviation from the SM prediction in the CP asymmetry in mixing of $\Bs$ meson. 
It requires a verification by the independent measurements at LHC, but regardless the outcome
of this cross check, the research pioneered at the Tevatron gives a strong boost
to further studies in this direction. The list of important results may be extended to many other 
excellent measurements
performed at the Tevatron. Regrettably, due to a luck of space, some of them are omitted from this review. 
But probably the most important achievement of the Tevatron experiments is a clear
and indubitable demonstration of a possibility and high value of performing $B$ physics
at hadron collider, which provides an important support for extending this program at the LHC.

\section*{Acknowledgements}
All results presented in this review are obtained by the common efforts of many people of the
CDF and D\O\ collaboration, and I want to expresses my gratitude to all my colleagues from both
experiments. Without their relentless work neither of these result was possible.
I also want to thank the scientists of the accelerating division of the Fermilab laboratory,
who provided a stable work of the Tevatron collider, breaking many records of performance,
which was crucial for obtaining all results presented here.

%\begin{thebibliography}{000} %for 3 digits

\end{document}